%% file: main.tex
\documentclass{egpubl}
\usepackage{egsgp2024}
 
\SpecialIssuePaper         

\CGFccby

\usepackage[T1]{fontenc}
\usepackage{dfadobe}  

\usepackage{cite}  
\BibtexOrBiblatex
\electronicVersion
\PrintedOrElectronic
\ifpdf \usepackage[pdftex]{graphicx} \pdfcompresslevel=9
\else \usepackage[dvips]{graphicx} \fi

\usepackage{egweblnk} 

\usepackage[dvipsnames, table]{xcolor}
\usepackage{mathtools}
\usepackage{amsfonts}
\usepackage[inline]{enumitem}
\usepackage{algorithm}
\usepackage{algpseudocode}
\usepackage{nicefrac}
\usepackage{overpic}
\usepackage{tikz}
\usepackage{subcaption}
\usepackage{booktabs}
\usepackage{xspace}
\usepackage{wrapfig}

\usepackage{tabularx} 
\newcolumntype{Y}{>{\centering\arraybackslash}X}

\usepackage{amsthm}

\usepackage[dvipsnames]{xcolor}

\definecolor{revisedcolor}{rgb}{0.85,0,0}
\newcommand{\revised}[1]{#1}

\usepackage{cleveref}

\def\Reals{\mathbb{R}}
\newcommand{\Realsn}[1]{\Reals^{#1}}
\newcommand{\manifold}[1]{\mathcal{#1}}
\newcommand{\mesh}[1]{\Hat{\manifold{#1}}}
\def\M{\manifold{M}}
\def\DiscM{\mesh{M}}
\def\C{\manifold{C}}
\def\DiscC{\mesh{C}}

\def\F{\manifold{F}}

\newcommand{\lfs}[1]{\mathrm{lfs}({#1})}
\newcommand{\glfs}[1]{\mathrm{ilfs}({#1})}
\newcommand{\reach}[1]{\mathrm{reach}({#1})}
\newcommand{\greach}[1]{\mathrm{ireach}({#1})}
\newcommand{\rball}[2]{\manifold{B}_{{#1}, {#2}}}
\newcommand{\closure}[1]{\mathrm{cl}{#1}}
\newcommand{\se}[1]{SE({#1})}
\newcommand{\so}[1]{SO({#1})}

\def\BigOSym{\mathcal{O}}
\newcommand{\BigO}[1]{\BigOSym\left({#1}\right)}

\newcommand{\mymat}[1]{\mathbf{#1}}

\DeclareMathOperator*{\argmin}{arg\,min}

\newtheorem{thm}{Theorem}
\newtheorem{dfn}{Definition}
\newtheorem{prp}{Proposition}
\newtheorem{lem}{Lemma}
\newtheorem{cor}{Corollary}

\newtheorem{innercustomthm}{Theorem}
\newenvironment{customthm}[1]
  {\renewcommand\theinnercustomthm{#1}\innercustomthm}
  {\endinnercustomthm}
  \newtheorem{innercustomprp}{Proposition}
\newenvironment{customprp}[1]
  {\renewcommand\theinnercustomprp{#1}\innercustomprp}
  {\endinnercustomprp}
  \newtheorem{innercustomlem}{Lemma}
\newenvironment{customlem}[1]
  {\renewcommand\theinnercustomlem{#1}\innercustomlem}
  {\endinnercustomlem}
  \newtheorem{innercustomcor}{Corollary}
\newenvironment{customcor}[1]
  {\renewcommand\theinnercustomcor{#1}\innercustomcor}
  {\endinnercustomcor}

\def\eg{\emph{e.g.}\xspace}
\def\ie{\emph{i.e.}\xspace}
\def\etal{\emph{et al.}\xspace}

\title[Reconstructing Curves from Sparse Samples on Riemannian Manifolds]%
    {Reconstructing Curves from Sparse Samples on Riemannian Manifolds}

\author[D. Marin et al.]%
{\parbox{\textwidth}{
    \centering 
        D. Marin$^{1}$\orcid{0000-0002-8812-9719}, 
        F. Maggioli$^{2}$\orcid{0000-0001-8008-8468},
        S. Melzi$^{2}$\orcid{0000-0003-2790-9591},
        S. Ohrhallinger$^{1}$\orcid{0000-0002-2526-7700},
        and
        M. Wimmer$^{1}$\orcid{0000-0002-9370-2663}
        }
        \\
{\parbox{\textwidth}{\centering $^1$TU Wien, Vienna, Austria\\
         $^2$University of Milano-Bicocca, Milan, Italy
       }
}
}

\begin{document}

\teaser{
 \centering
 \includegraphics[width=\linewidth]{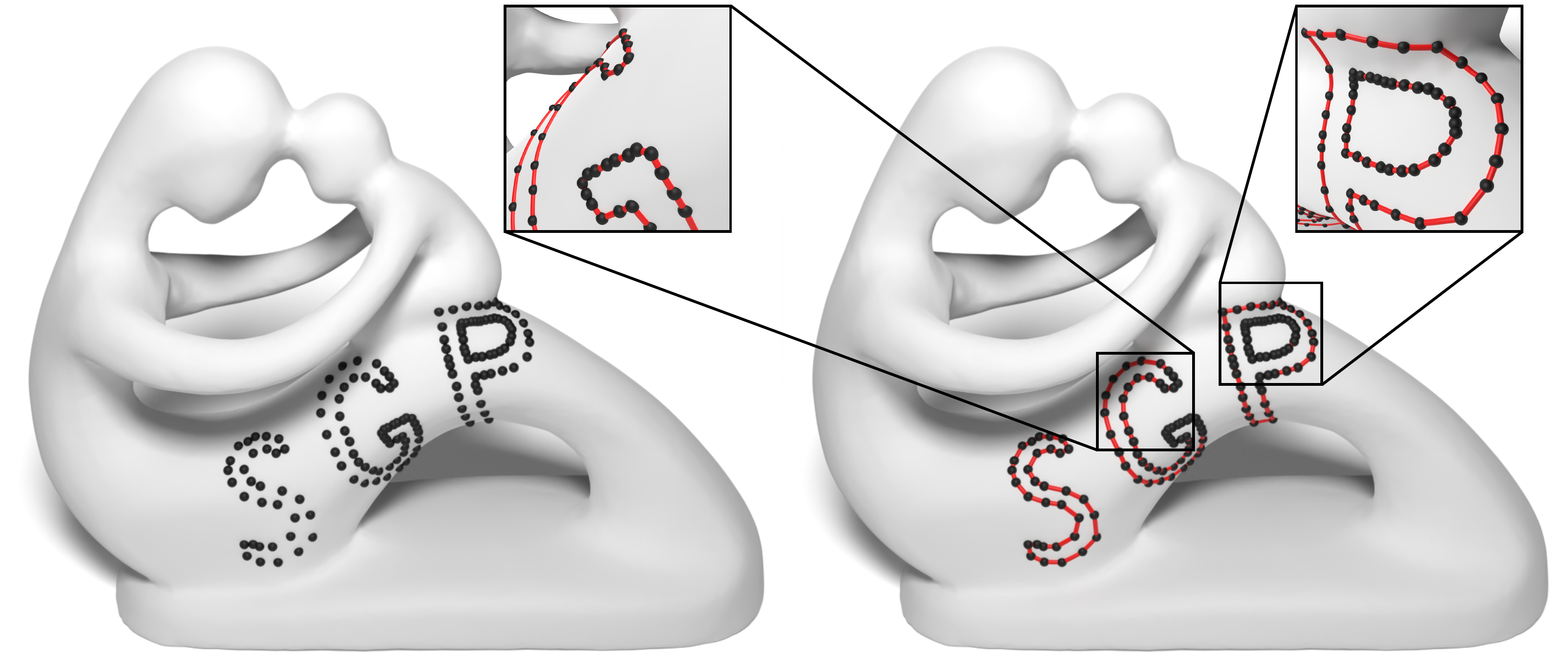}
  \caption{Reconstruction of multiple curves on the fertility mesh. The samples (left) are denser where the local feature size is small - around the serif of the \textit{G} for example, matching our sampling condition. Our method is able to reconstruct multiple closed curves at the same time (right), with sharp features and close sheets.}
\label{fig:teaser}
}

\maketitle
\begin{abstract}

\input{00-abstract/main}
\end{abstract}  
\section{Introduction}\label{sec:ïntro}

\input{01-intro/main}

\section{Related work}\label{sec:relwork}
\input{01.1-relwork/main}

\section{Background and notations}\label{sec:background}

\input{02-background/main}

\section{Method}\label{sec:method}

\input{03-method/main}

\section{Applications}\label{sec:results}

\input{04-results/main}

\section{Conclusions}\label{sec:conclusions}

\input{05-conclusions/main}

\section*{Acknowledgements}
This work has been partially funded by the Austrian Science Fund (FWF) project no. P32418-N31 and by the Wiener Wissenschafts-, Forschungs- und Technologiefonds (WWTF) project ICT19-009, as well as by the Italian Ministry of Education, Universities and Research under the grant \textit{Dipartimenti di Eccellenza 2023-2027} of the Department of Informatics, Systems and Communication of the University of Milano-Bicocca and by the PRIN 2022 project \textit{GEOPRIDE Geometric primitive fitting on 3D data for geometric analysis and 3D shapes}.
We acknowledge the support of NVIDIA Corporation with the RTX A5000 GPUs granted through the Academic Hardware Grant Program to the University of Milano-Bicocca for the project \textit{Learned representations for implicit binary operations on real-world 2D-3D data}.


\bibliographystyle{eg-alpha-doi} 
\bibliography{biblio}       


\appendix

\section{Proofs}\label{sec:proofs}

\input{a0-proofs/main}

\end{document}

%% file: 00-abstract/main.tex

Reconstructing 2D curves from sample points has long been a critical challenge in computer graphics, finding essential applications in vector graphics. The design and editing of curves on surfaces has only recently begun to receive attention, primarily relying on human assistance, and where not, limited by very strict sampling conditions. In this work, we formally improve on the state-of-the-art requirements and introduce an innovative algorithm capable of reconstructing closed curves directly on surfaces from a given sparse set of sample points. We extend and adapt a state-of-the-art planar curve reconstruction method to the realm of surfaces while dealing with the challenges arising from working on non-Euclidean domains. We demonstrate the robustness of our method by reconstructing multiple curves on various surface meshes. We explore novel potential applications of our approach, allowing for automated reconstruction of curves on Riemannian manifolds.

\ccsdesc[500]{Mathematics of computing~Paths and connectivity problems}
\ccsdesc[300]{Mathematics of computing~Graph algorithms}
\ccsdesc[300]{Computing methodologies~Mesh geometry models}

\printccsdesc

%% file: 01-intro/main.tex
Vector graphics represents an important research area in computer graphics, and it is widely applied in many fields, spanning from design and art to engineering. One of the important reasons for its success is the ability to generate infinite resolution smooth complex visualizations with relative ease while requiring only little input geometry. 

Recently, an increasing interest has been devoted to moving 2D vector graphics onto surfaces, trying to address certain issues stemming from texturing methods. Texturing is a well-established approach for defining patterns and decorations on surfaces, but it generally relies on finite-resolution images and parameterization. The latter is not always available or could be difficult or expensive to define. Procedural textures try to overcome these problems, defining patterns via mathematical functions and algorithms. However, they generally rely on multi-dimensional noise functions that are then sampled on the surface~\cite{ebert:2003:texturing,hart:2001:perlin,maggioli:2022:newton}. These algorithms are usually agnostic of the underlying geometric properties and can incur high computation times.

Besides works generalizing sample-based texture synthesis to triangular meshes~\cite{wei:2001:synthesis,turk:2001:synthesis}, only in recent years some solutions have been proposed which try to leverage properties of non-Euclidean metric spaces and define patterns directly on surfaces, either via recursive structures~\cite{nazzaro:2021:geotangle} or simulated behaviors~\cite{maggioli:2022:momas}. Another avenue of development for texture synthesis is represented by neural networks that generate a textured mesh in the style of an input image~\cite{shailesh:2023:styletransfer}. 

Despite the existence of various solutions for decorating surfaces, the problem of constructing lines and curves on discrete manifolds has not been addressed satisfactorily yet. Defining curves and shapes directly on surfaces is innovative for design applications~\cite{poerner:2018:adidas}, and has a relevant impact in the processing of archaeological data~\cite{kolomenkin:2008:curvesillustrations,gilboa:2013:curvearcheo} by extracting specific decorations from the models. However, little research has been devoted to improving the definition and the reconstruction of curves on discrete surfaces, besides efforts to generalize B\'{e}zier curves~\cite{mancinelli:2021:bsurf}. Furthermore, the existing techniques are generally centered around human interaction, as they are designed to be tools for artists and end users, and even the most recent solutions require an ordered sequence of samples~\cite{mancinelli:2023:centermass}. To the best of our knowledge, Shah \etal~\cite{shah:2013:curve} provided the first and only solution for dealing with curve reconstruction on Riemannian manifolds. Still, their proposed method can only deal with dense uniform samplings and only guarantees to reconstruct curves in limited settings. By generalizing state-of-the-art theoretical results and algorithmic solutions for planar curve reconstruction to arbitrary manifold domains, we introduce a more robust method that, given an unordered collection of points over a Riemannian manifold that follows our relaxed sampling conditions, always produces an ordered sequence identifying a closed curve on the surface (see \Cref{fig:teaser}).

Our contributions are summarized in the following:
\begin{itemize}[topsep=0pt,noitemsep]
    \item we propose a solution that extends existing state-of-the-art theory and techniques for 2D curve reconstruction to manifold domains, overcoming the challenges arising from translating the problem into non-Euclidean spaces;
    
    \item we improve the state-of-the-art sampling conditions for curve reconstruction on Riemannian manifolds~\cite{shah:2013:curve}, allowing for sparser sampling for curve reconstruction on manifold domains;

    \item we perform a qualitative study of our method on real-world data coming from established applications, where the previous solution fails due to the sparsity of the samples.

\end{itemize}

%% file: 01.1-relwork/main.tex
Reconstructing curves from samples in a non-Euclidean space extends the classical problem of planar curve reconstruction to more complex spaces, drawing inspiration from tangent domains: surface design and texturing using on-surface elements. This section will provide an overview of various techniques that deal with each of these fields individually and explain how they relate to our work.

\textbf{Planar curve reconstruction}, where the input samples and their respective reconstructed curve live in $\mathbb{R}^2$, is dealt with by numerous methods. They are usually divided into two main categories: implicit (methods that approximate the inside/outside of the shape and recreate the boundary based on the division between the two)~\cite{hoppe:1992:surfrecon,kazhdan:2006:poisson} and explicit (interpolatory methods that create some ordering among the sample points). We will focus on the explicit reconstruction methods, as they are the most relevant to our work. However, most of these methods are limited to planar curves, with some of them extended to reconstructing surfaces as well. Some of the presented methods can reconstruct curves in $\mathbb{R}^3$ but the samples do not live on a manifold.

To interpolate the input sample points, most methods compute a graph on the input and use a subset of the edges as the final reconstruction. One of the most commonly used types of graphs is the Delaunay triangulation, due to its geometrical properties and theoretical guarantees of including the reconstruction subject to sampling density~\cite{amenta:1998:crust}. Starting from the Delaunay triangulation, Amenta \etal filter the edges whose proximity is empty of samples and of Voronoi vertices. This approach has been improved to take into account Voronoi poles - Voronoi vertices corresponding to long, skinny Voronoi cells~\cite{amenta:2001:powercrust}.

Using the same Delaunay starting base, a greedy procedure is used to pick a seed vertex and find the nearest neighbors until the endpoints are close enough to be connected or until all points have been connected~\cite{Parakkat:2016:crawl}. Similarly, various criteria have been devised for keeping a Delaunay edge between two points: the new neighbor has to be situated in the half-plane (defined using the normal at the current sample) opposite to the previous connection~\cite{dey:1999:nncrust}, or in the opposite half-plane defined using the bisector of the previous edge~\cite{ohrhallinger:2016:hnncrust}. Another set of criteria used to filter the triangulation is based on leveraging the Voronoi poles to approximate the normals for the half-plane computation, considering the angle and the ratio between the current edge and its Voronoi counterpart~\cite{dey:2002:gathang}. Various other methods build on the Delaunay triangulation as a starting base for curve reconstruction~\cite{ohrhallinger:2013:connect2d, marin:2022:sigdt}, and a comprehensive survey on multiple implicit methods can be found here~\cite{ohrhallinger:2021:benchmark}. Our method lifts the reconstruction of curves from planar surfaces to manifold domains of any dimension.

\pagebreak
\textbf{Curve reconstruction on Riemannian manifolds} has been approached as an extension of planar metrics and sampling conditions to Riemannian manifolds~\cite{shah:2013:curve}. However, the classical definition of the medial axis does not hold on surfaces if the medial axis is constrained to live on the surface as well. The authors introduce a new sampling criterion, based on the minimum value between the distance to the medial axis and the injectivity radius. We introduce the formal definitions of these concepts in \Cref{sec:background} and we relax their dense sampling requirements in \Cref{sec:method} to allow for reconstruction with fewer samples. To the best of our knowledge, this is the only work that extends the problem of curve reconstruction to Riemannian manifolds.


\textbf{Vector graphics} on planar surfaces have been thoroughly researched and are being used in multiple tools~\cite{Inkscape, adobeillustrator}. Recently, editing and importing curves on surfaces have received interest in the graphics field, partly due to the improvement in computing geodesic paths efficiently~\cite{sharp:2020:flipout}. Users can interact with designs directly on the mesh, by either editing splines on a 2D local projection of the surface~\cite{poerner:2018:adidas}, which is usually prone to artifacts due to the projection procedure, as explored in \cite{yuksel:2019:texturemapping}, or editing splines directly on the surfaces, bypassing the projection artifacts, by using geodesic metrics on the surface~\cite{nazzaro:2021:geotangle, mancinelli:2021:bsurf}. However, these editing methods require user input or a predefined ordering of the samples to construct the curves on the surface, which is the missing link we are providing. Hence, our method also provides a building block for further editing splines on surfaces.

%% file: 02-background/main.tex
Informally speaking, a $d$-dimensional manifold is a collection of pieces of the $d$-dimensional hyperspace that are deformed and glued together to form a continuous smooth domain. Such domain can be embedded in a higher dimensional space (like in the case of 2-dimensional surfaces embedded in 3D space), or exist on their own (like the 3D rotation group $\mathrm{SO}(3)$~\cite{hall:2013:liegroups}). For a formal and complete definition of Riemannian manifolds, we refer to the books by Morita and Do Carmo~\cite{morita:2001:geometry,docarmo:2016:geometry}.

To refer to any such $d$-dimensional manifold (including continuous curves) we use calligraphic letters (\ie, $\M$, $\C$), and we denote the geodesic distance function over a manifold $\M$ with $d_{\M}$. Furthermore, we refer to any discrete representation of a manifold $\M$ with the symbol $\DiscM$.

Specifically, we represent discrete $2$-dimensional surfaces as triplets $\DiscM = (V, E, T)$, where $V$ is a set of vertices, $T$ is a set of oriented triangles among vertices, and $E$ is a set of unordered edges induced by the triangles in $T$. We also represent discrete curves (\ie, 1-dimensional manifolds) as tuples $\DiscC = (V, E)$, where $V$ is a set of vertices and $E$ is a set of edges between vertices.
Throughout this work, we assume the manifoldness of discrete surfaces and curves~\cite{crane:2013:discretegeometry}, and we assume to work with curves made by a single component and forming a closed non-self-intersecting loop.

We also recall the definitions of \emph{medial axis} and \emph{local feature size} (see \Cref{fig:medial-axis-rho2d}), which we will use for distinguishing between different curve sampling strategies.
\begin{dfn}[\hspace{1sp}\cite{blum:1967:medialaxis}]\label{dfn:medialaxis}
    Let $\C$ be a smooth curve embedded in a metric space $(\M, d_{\M})$. The \emph{medial axis} $\Gamma(\C)$ of $\C$ is the closure of the set of points in $\M$ that have two or more closest points in $\C$.
\end{dfn}
\begin{dfn}[\hspace{1sp}\cite{ruppert:1993:lfs}]\label{dfn:lfs}
    Let $p \in \C$ be a point on the curve. The \emph{local feature size} $\lfs{p}$ of $p$ is the minimum distance from $p$ to the medial axis of $\C$.
    \begin{equation}
        \lfs{p}
        =
        \min_{q \in \Gamma(\C)} d_{\M}(p, q)\,.
    \end{equation}
\end{dfn}

\subsection{Curve sampling}\label{sec:background:curverecon}

\input{02-background/02-curverecon/main}

\subsection{Differential geometry}\label{sec:background:diffgeo}

\input{02-background/01-diffgeo/main}

%% file: 02-background/02-curverecon/main.tex

In this section, we repeat definitions related to sampling planar curves and conclude with our own definitions that extend to manifolds.

Given a smooth curve $\C$ and a set of samples $S \subset \C$, reconstructing the curve is the process of constructing a discrete curve $\DiscC = (S, E)$ such that an edge $e = (s_i, s_j)$ is in $E$ if and only if $s_i$ and $s_j$ are consecutive samples along the curve $\C$. 



\begin{figure}
    \centering\includegraphics[width=\linewidth]{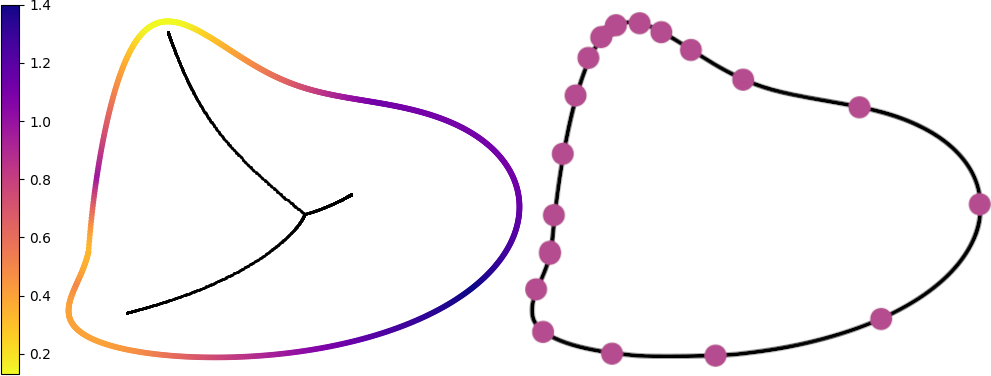}
    \caption{Left: the local feature size at each point of the curve represents its distance to the medial axis (in black). Right: a $\rho$-sampling of the curve with $\rho = 1.0$ shows that the samples are denser where the medial axis is closer to the curve.}
    \label{fig:medial-axis-rho2d}
\end{figure}

\begin{figure}[t]
    \centering
    \includegraphics[width=\columnwidth]{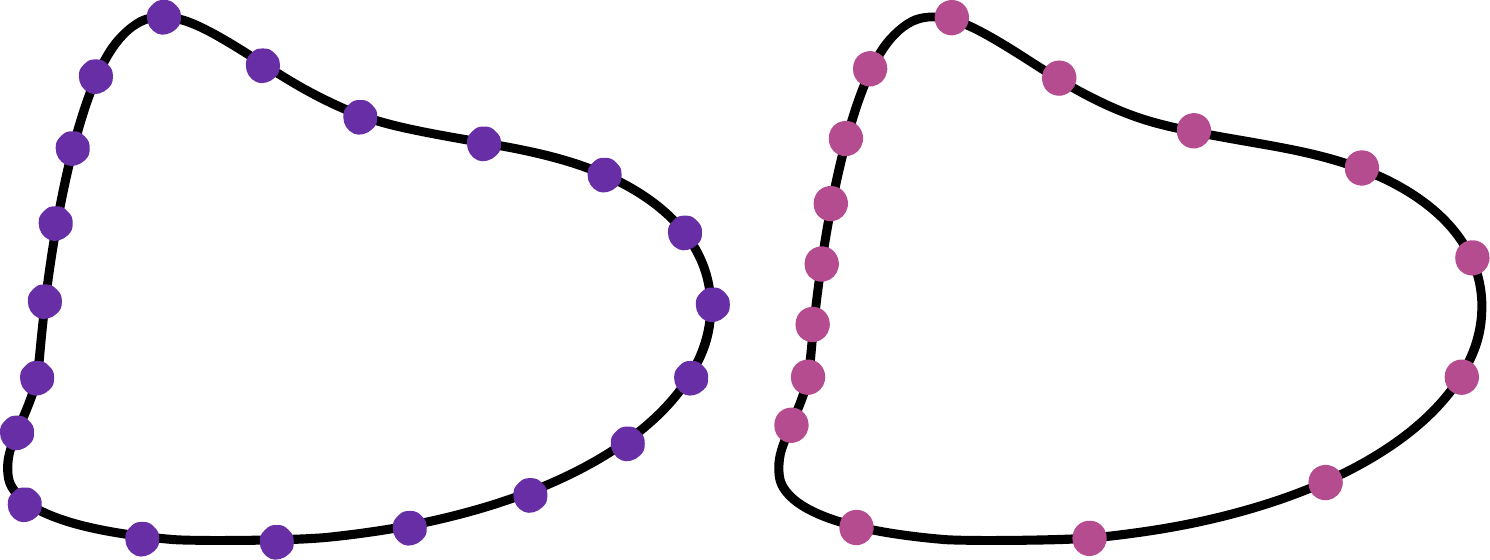}
    \caption{Difference between uniform sampling (left) and non-uniform sampling (right) with non-uniformity ratio $u=1.5$.}
    \label{fig:uniform-nonuniform-diff2d}
\end{figure}


One technique to evaluate the sampling quality is by using the $\rho$ value, which was introduced by Ohrhallinger \etal~\cite{ohrhallinger:2016:hnncrust} and relies on the notion of \emph{reach}.

\begin{dfn}[\hspace{1sp}\cite{federer:1959:curvature}]\label{dfn:reach2d}
    The \emph{reach} of an interval $I \subset \C$ is the minimum local feature size among points in the interval:
    \begin{equation}
        \reach{I}
        =
        \inf_{p \in I}\ \lfs{p} \,.
    \end{equation}
\end{dfn}
\begin{dfn}[\hspace{1sp}\cite{ohrhallinger:2016:hnncrust}]\label{dfn:rhosampling}
    A smooth curve $\C$ is \emph{$\rho$-sampled} by a sample set $S \subset \C$ if every point $p \in \C$ of the curve is closer to a sample than a $\rho$-fraction of the reach of the interval $I = [s_0, s_1]$ of consecutive samples containing it:
    \begin{equation}
        \forall p \in I,\ 
        \min_{s \in \{ s_0, s_1 \}}\ d_{\M}(p, s) < \rho\ \reach{I} \,.
    \end{equation}
\end{dfn}

In \Cref{fig:medial-axis-rho2d}, we show how the $\rho$-sampling depends on the local feature size, requiring a more dense sampling where the medial axis is closer to the curve and allowing a more sparse sampling where the local feature size is larger.

A different category of sampling conditions enforces a specific distance between consecutive samples, unrelated to the medial axis. Hence, we conclude our collection of planar sampling conditions by distinguishing between uniform and non-uniform sampling, examples of which can be observed in \Cref{fig:uniform-nonuniform-diff2d}.
\begin{dfn}\label{dfn:uniformsampling}
    A smooth curve $\C$ is \emph{uniformly $\vartheta$-sampled} by a sample set $S \subset \C$ if for any two consecutive samples $s_0, s_1 \in S$, it holds $d_{\M}(s_0, s_1) < \vartheta$.
\end{dfn}
\begin{dfn}\label{dfn:nonuniformsampling}
    A smooth curve $\C$ is \emph{non-uniformly sampled with a non-uniformity ratio $u \in \Realsn{+}$} by a sample set $S \subset \C$ if for any three consecutive samples $s_0, s_1, s_2 \in S$, it holds that $u_{s_1} < u$, where the $u_{s_1}$ is defined as:
    \begin{equation}
        u_{s_1}
        =
        \frac{
            \max\left( d_{\M}(s_0, s_1), d_{\M}(s_1, s_2) \right)
        }{
            \min\left( d_{\M}(s_0, s_1), d_{\M}(s_1, s_2) \right)
        }\,.
    \end{equation}
\end{dfn}

%% file: 02-background/01-diffgeo/main.tex
We now lift the definitions required for the local feature size from the plane to manifolds, to be able to introduce sampling conditions on manifolds in the subsequent sections.

Euclidean disks are widely used in relation to planar curve reconstruction, and they have a straightforward generalization to arbitrary metric spaces.
\begin{dfn}\label{dfn:rball}
    Let $(\M, d_{\M})$ be a metric space, $p \in \M$ a point on $\M$, and $r \in \Reals$. The \emph{$r$-ball centered at $p$} is the region $\rball{p}{r}$ of points at a distance less than $r$ from $p$:
    \begin{equation}
        \rball{p}{r}
        =
        \left\lbrace
        x \in \M\ 
        :\ 
        d_{\M}(p, x) < r
        \right\rbrace
        \,.
    \end{equation}
    The closure $\closure{\rball{p}{r}}$ of the $r$-ball also includes its boundary:
    \begin{equation}
        \partial\rball{p}{r}
        =
        \left\lbrace
        x \in \M\ 
        :\ 
        d_{\M}(p, x) = r
        \right\rbrace
        \,.
    \end{equation}
\end{dfn}

\begin{figure}
    \centering
    \includegraphics[width=0.5\columnwidth]{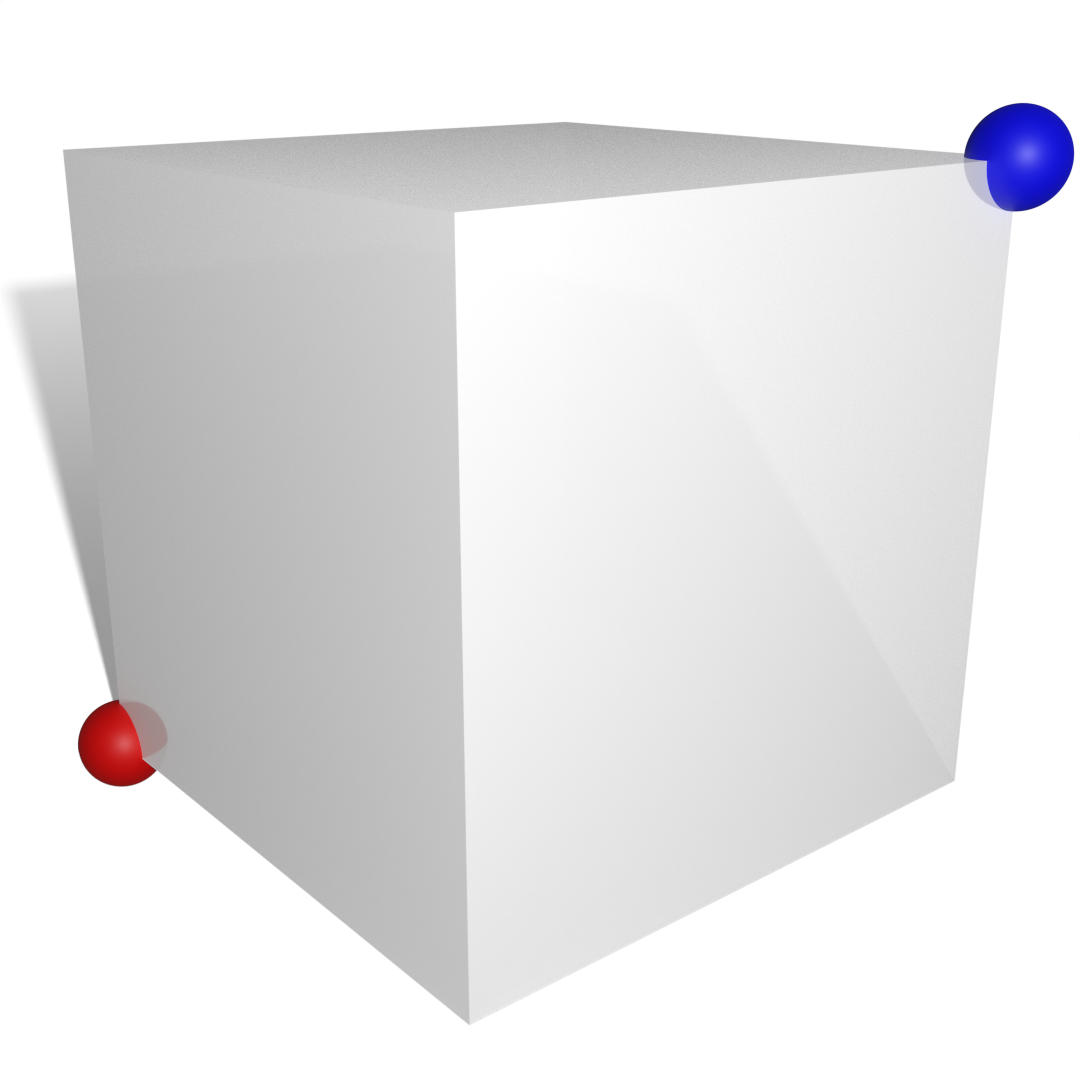}%
    \includegraphics[width=0.5\columnwidth]{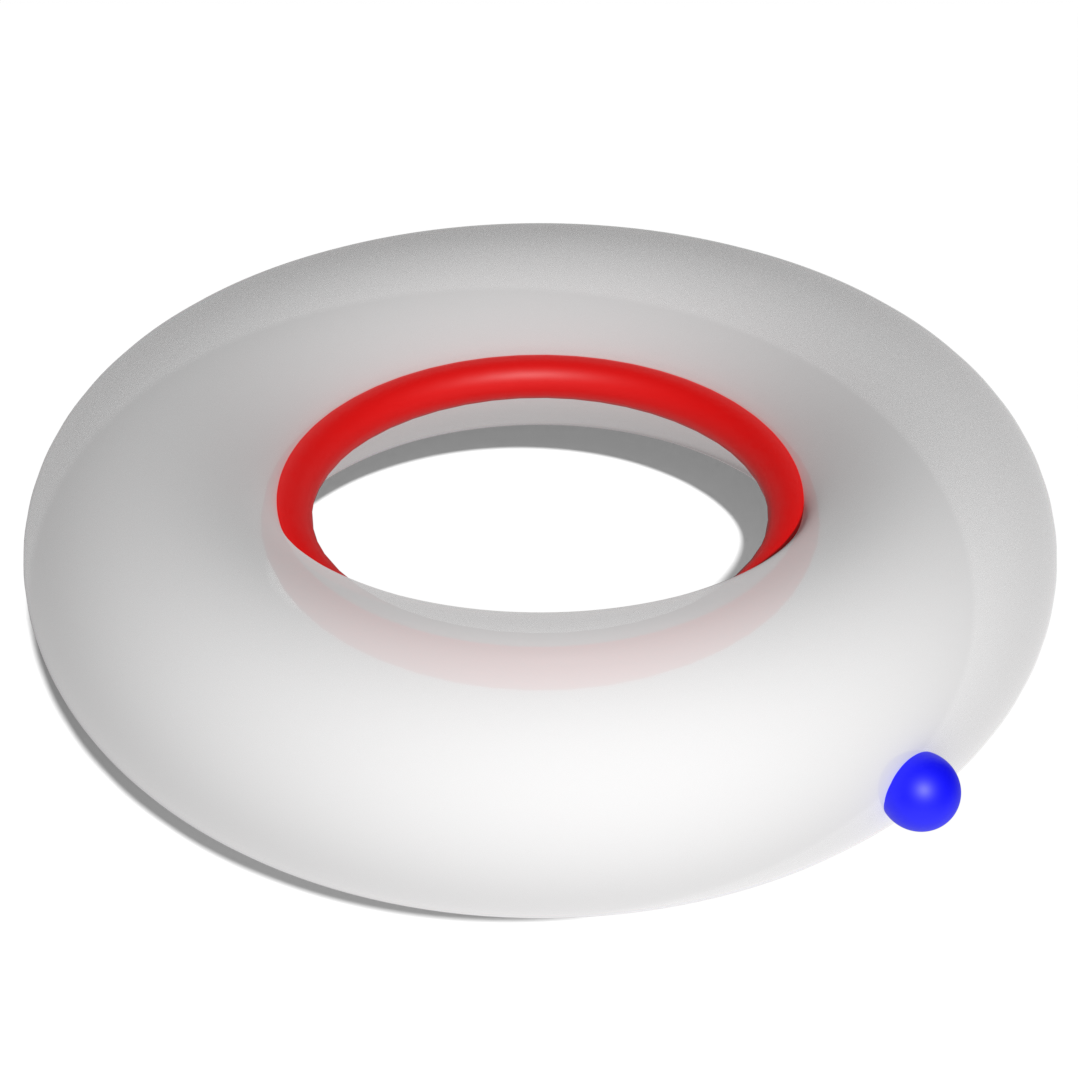}
    \caption{Examples of cut locus (in red) given a point (in blue) on different surfaces. The cut locus can be a single point (left) or an entire curve (right).}
    \label{fig:cut-locus-sample}
\end{figure}

The work of Shah \etal~\cite{shah:2013:curve} identifies the limitations of the local feature size in non-Euclidean metric spaces, and exploits the notions of \emph{cut locus} and \emph{injectivity radius} to overcome these issues.
\revised{Informally speaking, the cut locus of a point $p$ on the manifold $\M$ can be seen as the set of all points $q$ such that there are at least two distinct minimizing geodesics from $p$ to $q$ -- see \Cref{fig:cut-locus-sample}.}
\begin{dfn}\label{dfn:cutlocus}
    Let $\M$ be a $d$-dimensional Riemannian manifold, possibly with boundary $\partial\M$, and equipped with a connection defining an exponential map $\exp_p : T_p(\M) \to \M$ at every point $p \in \M$. The \emph{cut locus of $p$ in the tangent space $T_p(\M)$} is defined as the set $C_{T(\M)}(p)$ of all vectors $v \in T_p(\M)$ such that the parametric curve $\exp_p(t v)$ is a minimizing geodesic for $t \in [0, 1]$ and not minimizing for $t > 1$.

    The \emph{cut locus of $p$ on the manifold $\M$} is the set $C_{\M}(p)$ of all points $q \in \M$ that are the image of a vector $v \in C_{T(\M)}$ under the exponential map.
\end{dfn}
\revised{Intuitively, the injectivity radius is the maximum size of an $r$-ball around a point that preserves injectivity when mapped the Euclidean space -- refer to \Cref{thm:injradiusexpmap} and \Cref{fig:injrad-theorem}.}
\begin{dfn}[\hspace{1sp}\cite{docarmo:2016:geometry}]\label{dfn:injradius}
    Let $p \in \M$ be a point on the manifold $\M$ and $C_{\M}(p)$ be its cut locus. The \emph{injectivity radius of $p$} is
    \begin{equation}
        i_{\M}(p)
        =
        \inf_{q \in C_{\M}(p)} d_{\M}(p, q)\,.
    \end{equation}

    The \emph{injectivity radius of the manifold $\M$} is thus defined as
    \begin{equation}
        \mathfrak{I}_{\M}
        =
        \inf_{p \in \M} i_{\M}(p)\,.
    \end{equation}
\end{dfn}

By using these tools, Shah \etal~\cite{shah:2013:curve} show that curves can be reconstructed using a minimum spanning tree, if they are uniformly $\vartheta$-sampled, with $\vartheta < \min\left( \inf_{p \in \C} \lfs{p}, \mathfrak{I}_{\M} \right)$.

%% file: 03-method/main.tex
\begin{figure}[t]
    \centering
    \includegraphics[width=\columnwidth]{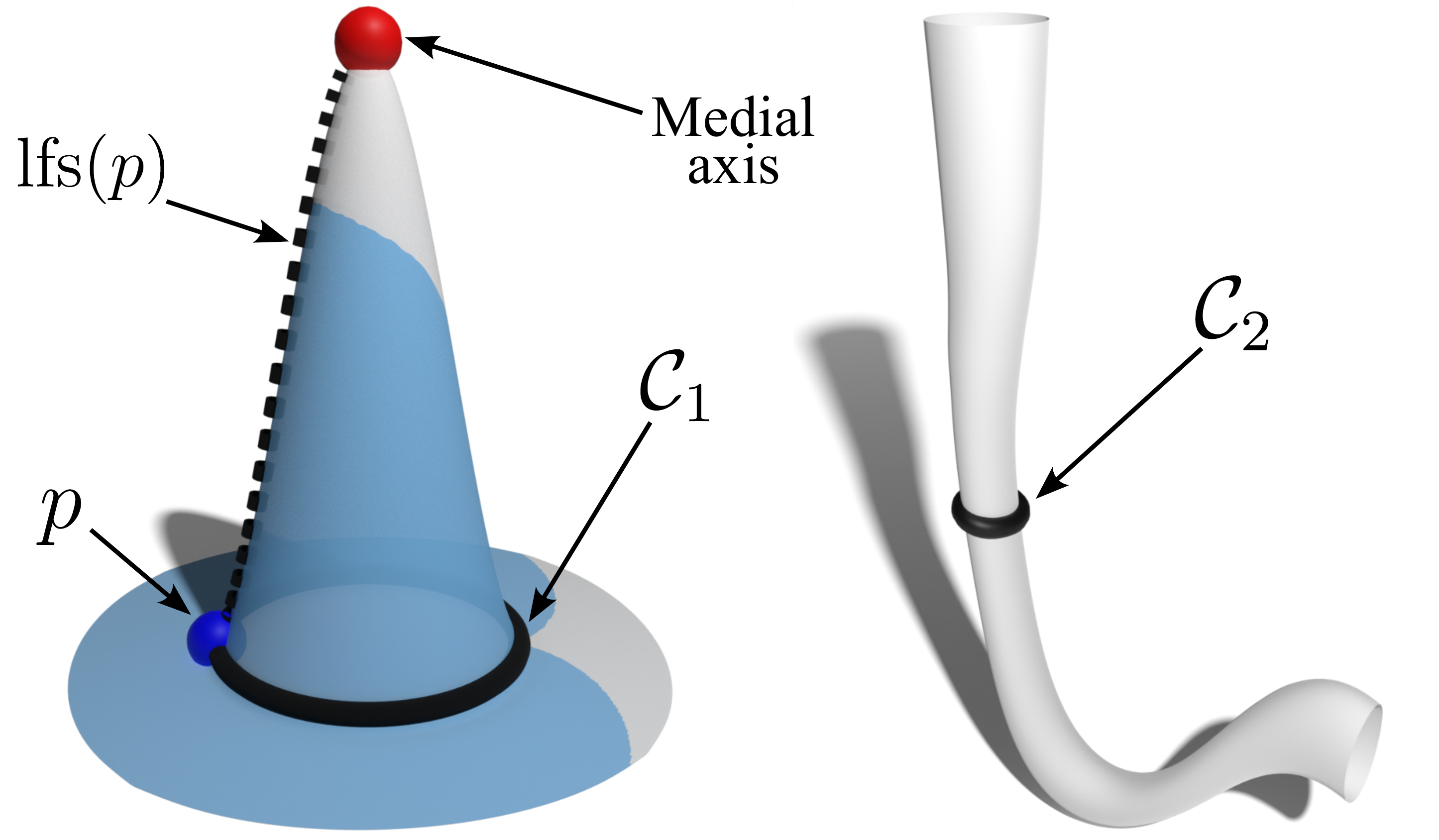}
    \caption{Two examples of curves on surfaces (solid black) where the local feature size exhibits undesired behaviors. Left: an $r$-ball (shaded in light blue) around $p$ that contains the entire curve $\C_1$, but no point of the medial axis. Right: a curve $\C_2$ on the surface with an empty medial axis (\ie, the local feature size is undefined).}
    \label{fig:lfs-fail}
\end{figure}

In this paper, we advance the task of reconstructing curves on manifolds, addressing curves that are sampled more sparsely and non-uniformly compared to the state-of-the-art.

The standard notion of local feature size is not suitable for non-Euclidean metric spaces. In \Cref{fig:lfs-fail} we show two examples of curves where the local feature size presents undesired behaviors. For instance, we can have $r$-balls that contain the entire curve without containing the medial axis, or even curves that do not have a medial axis, making it impossible to define the local feature size.

By taking inspiration from Shah \etal~\cite{shah:2013:curve}, we exploit the injectivity radius to strengthen the definition of $\rho$-sampling, making it suitable for manifold domains and showing that we can preserve its fundamental properties (\Cref{sec:method:sampling}). Then, we generalize a proximity graph successfully used for 2D curve reconstruction to arbitrary metric spaces, proving that analogous sampling conditions apply (\Cref{sec:method:sigdv}). And finally, we present an algorithm that relies on the properties of our graph to reconstruct curves on Riemannian manifolds (\Cref{sec:method:tsp}). The proofs of all our claims are provided in \Cref{sec:proofs}.

\subsection{Non-uniform sparse sampling on surfaces}\label{sec:method:sampling}

\input{03-method/01-sampling/main}

\subsection{SIGDV graph on manifolds}\label{sec:method:sigdv}

\input{03-method/02-sigdv/main}

\subsection{Curve reconstruction as a Traveling Salesman Problem}\label{sec:method:tsp}

\input{03-method/03-tsp/main}

%% file: 03-method/01-sampling/main.tex
The solution proposed by Shah \etal~\cite{shah:2013:curve} consists of bounding the local feature size to the injectivity radius of the entire surface. While this approach effectively solves the previous issues, it also makes the sampling very dense when the manifold contains very narrow sections or sharp features: the smallest injectivity radius of the surface affects the overall sampling conditions, even if the curve might not pass close to these regions. Instead, we use the local feature size and the reach by considering the injectivity radius of individual points.
\revised{In this way, we can define properties inside topologically flat regions and ensure that they behave similarly to a Euclidean space, avoiding cases where the topology of the $r$-ball introduces pathological issues -- see the left frame of \Cref{fig:lfs-fail}, where the local feature size is larger than the injectivity radius.}

\begin{figure}[t]
    \centering
    \includegraphics[trim={6cm 0 6cm 0},clip,width=0.49\columnwidth]{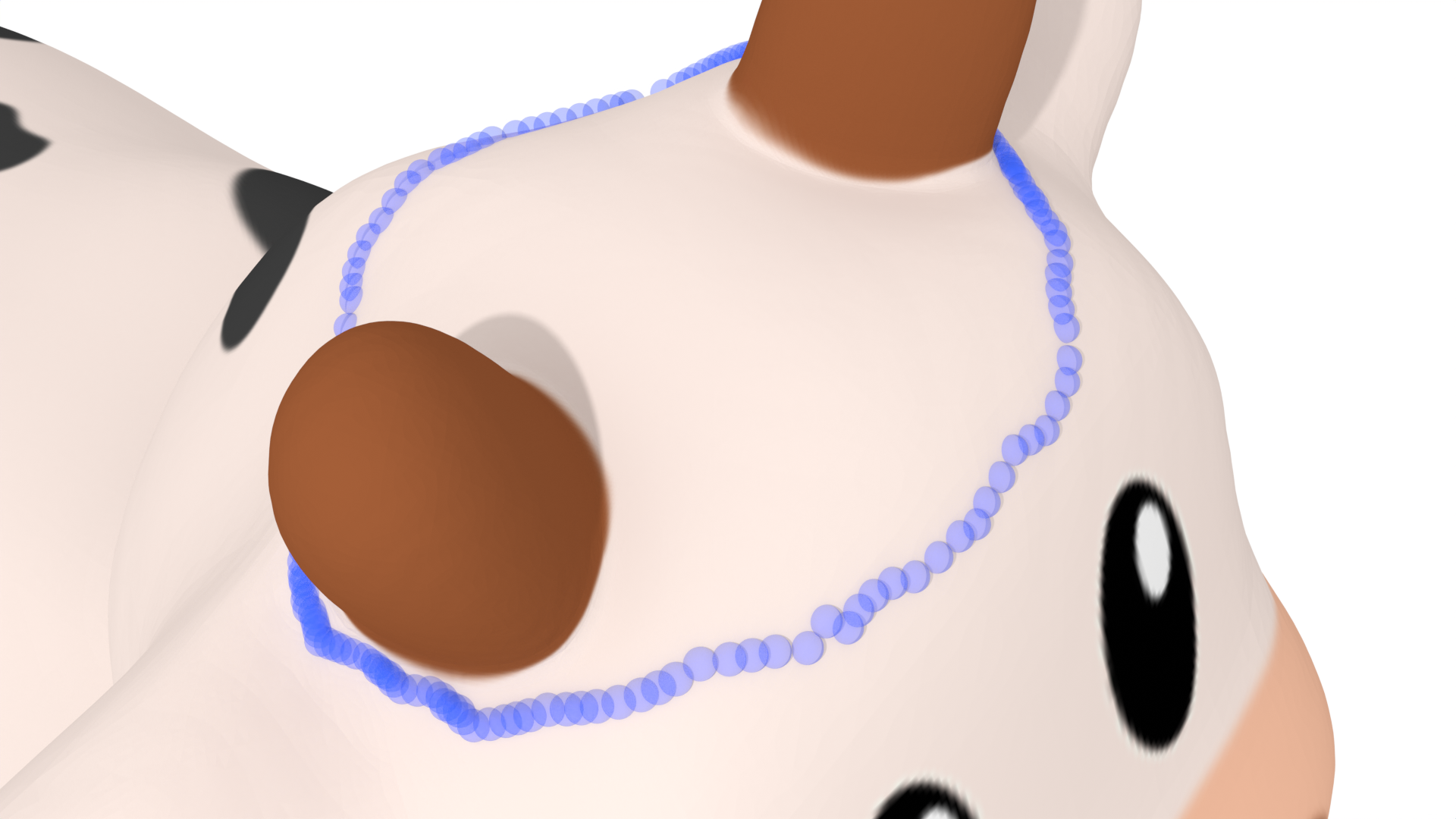}\hfill
    \includegraphics[trim={6cm 0 6cm 0},clip,width=0.49\columnwidth]{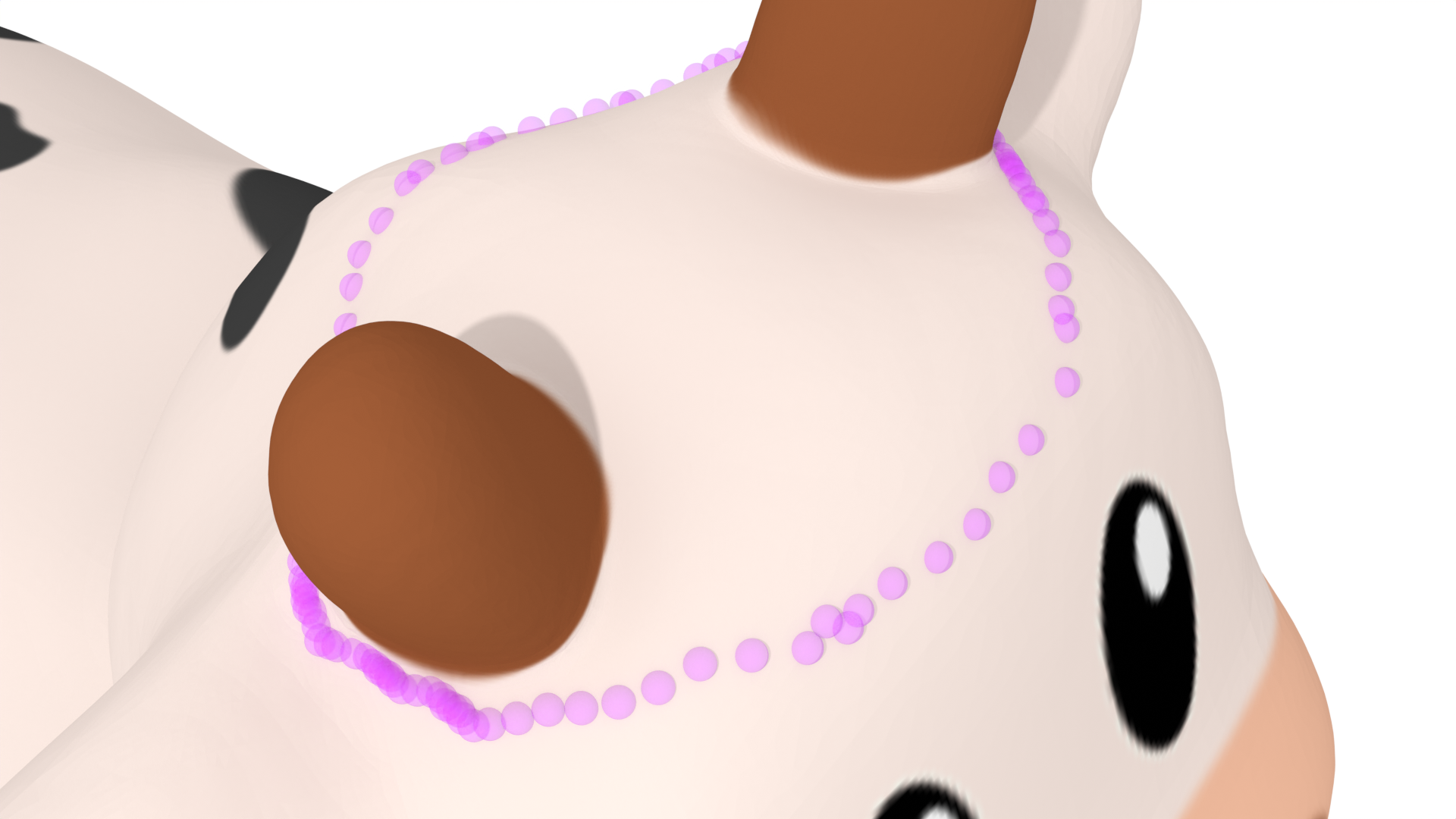}

    \caption{A comparison between a dense uniform sampling, with 281 samples, satisfying the conditions described by Shah \etal~\cite{shah:2013:curve} (on the left) and a non-uniform $\rho$-sampling scheme with only 124 samples (on the right).}
    
    
    \label{fig:sampling_diff}
\end{figure}

\begin{dfn}\label{dfn:geodesiclfs}
    Let $\C \subset \M$ be a closed curve on the manifold $\M$. For every point $p \in \C$, we define the \emph{injective local feature size} of $p$ as
    \begin{equation}
        \glfs{p}
        =
        \min\left( \lfs{p}, i_{\M}(p) \right)
        \,.
    \end{equation}

    For every interval $I \subset \C$ of the curve, we extend the \emph{injective reach} of $I$ as
    \begin{equation}
        \greach{I}
        =
        \min_{p \in I}\  \glfs{p}\,.
    \end{equation}
\end{dfn}
From \Cref{dfn:geodesiclfs}, the extension of $\rho$-sampling to its injective counterpart follows naturally. The example from \Cref{fig:sampling_diff} shows the difference between the uniform dense sampling required for the method from Shah \etal~\cite{shah:2013:curve} against a more relaxed non-uniform $\rho$-sampling scheme.

\revised{The sample sets in \Cref{fig:sampling_diff} were computed from a discrete, very dense initial curve by approximating the medial axis, and implicitly, the local feature size, using the Voronoi diagram~\cite{dey:2004:approximating}. We then employed a backtracking approach to extract a subset of samples that satisfy our sampling conditions. Generating an exact and minimal sampling remains an open problem.}

\paragraph*{Curve properties.}

\begin{figure}[t]
    \centering
    \includegraphics[width=0.5\columnwidth]{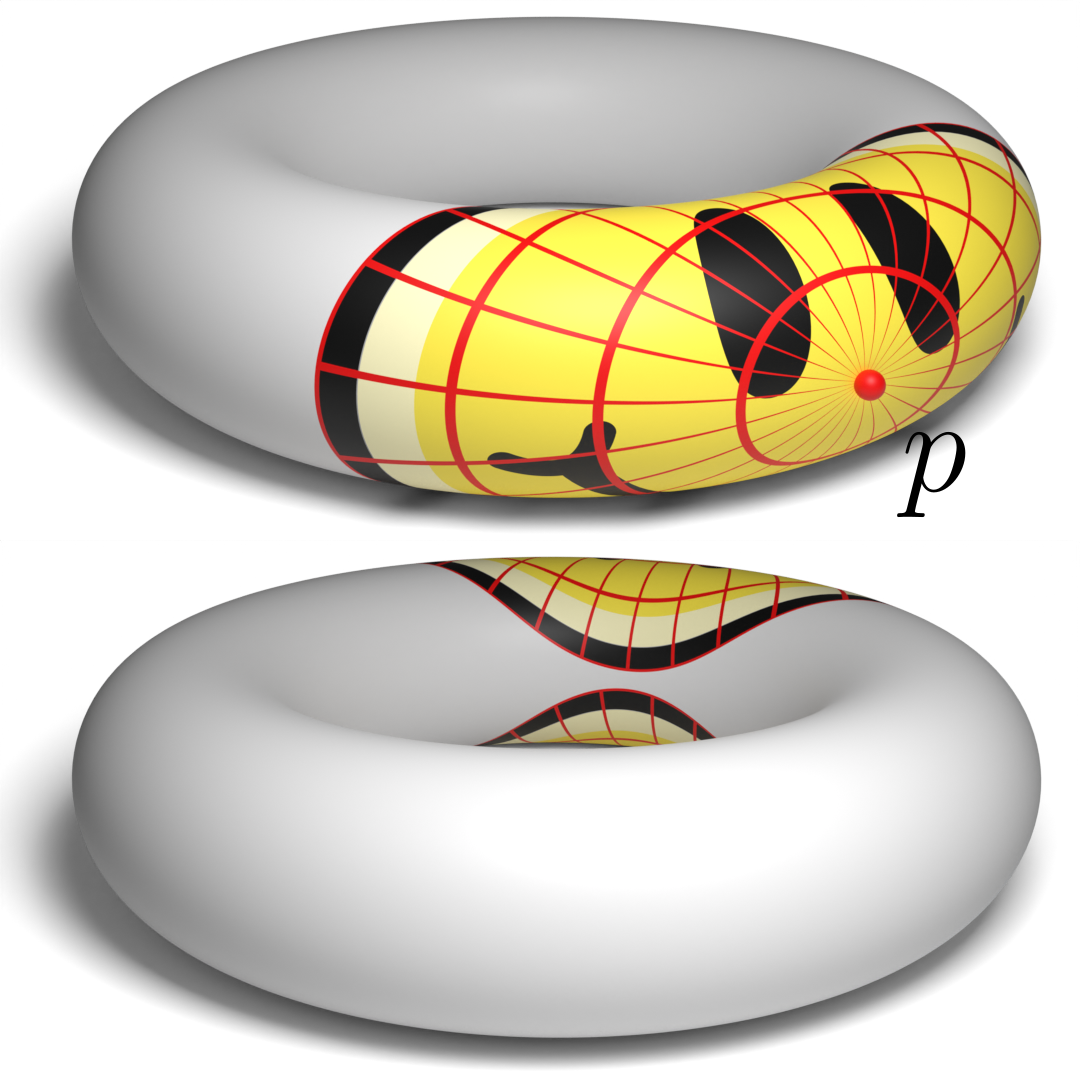}%
    \includegraphics[width=0.5\columnwidth]{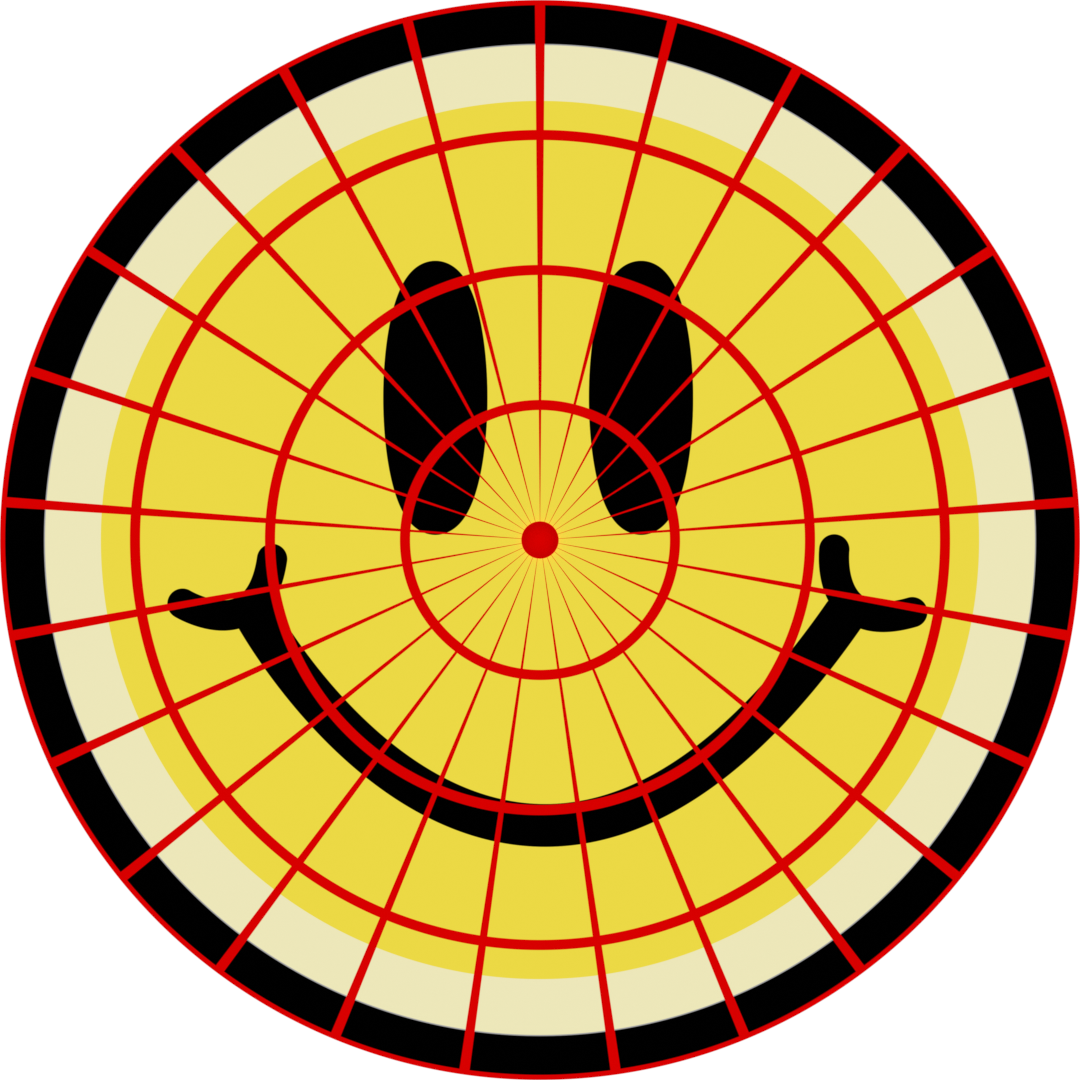}%
    \caption{Mapping of an $r$-ball around a point $p$ on a surface (left) to the Euclidean disk $\mathbb{B}^2$ (right). The mapping is possible because $r \leq i_{\M}(p)$.}
    \label{fig:injrad-theorem}
\end{figure}

We start from a classical result in differential geometry about the injectivity radius.
\begin{thm}[\hspace{1sp}\cite{klingenberg:1995:geometry}]\label{thm:injradiusexpmap}
    Let $\M$ be a $d$-dimensional Riemannian manifold. For every point $p \in \M$, the restriction of the exponential map $\exp_p$ to $U \subset T_p(\M)$, such that $\exp_p(U) = \rball{p}{r}$, $\rball{p}{r}$ is injective for all $r \leq i_{\M}(p)$ and it is a diffeomorphism onto its image.
\end{thm}
From this result, we can deduce that if $r \leq i_{\M}(p)$, then there exists a diffeomorphism between the $r$-ball $\rball{p}{r}$ centered at $p$ and the Euclidean $d$-dimensional ball $\mathbb{B}^{d}$, as we can see in the example depicted in \Cref{fig:injrad-theorem}. We can thus generalize the following important result that holds in 2D:

\begin{lem}[\hspace{1sp}\cite{amenta:1998:crust}]\label{lem:anytopodisk2d}
    Let $\C \subset \Realsn{2}$ be a closed curve in the plane. For every point $p \in \Realsn{2}$ and every $r \in \Realsn{+}$, if the Euclidean disk $D$ of radius $r$ and centered at $p$ contains at least two points of $\C$, then either $D\cap\C$ is a topological 1-disk, or $D$ contains a point of $\Gamma(\C)$, or both.
\end{lem}

For Riemannian manifolds of arbitrary dimensions we get:

\begin{lem}\label{lem:anytopodisk}
    Let $\C \subset \M$ be a closed curve on the manifold $\M$. For every point $p \in \M$ and every positive real value $r \leq i_{\M}(p)$, if the $r$-ball $\rball{p}{r}$ centered at $p$ contains at least two points of $\C$, then then either $\rball{p}{r}\cap\C$ is a topological 1-disk, or $\rball{p}{r}$ contains a point of $\Gamma(\C)$, or both.
\end{lem}
\begin{cor}\label{cor:curvetopodisk}
    For every point $p \in \C$, and for $r \leq \glfs{p}$, the ball $\rball{p}{r}$ intersects $\C$ in a topological 1-disk.
\end{cor}

\revised{%

\begin{figure}[t]
    \centering
    \includegraphics[width=\columnwidth]{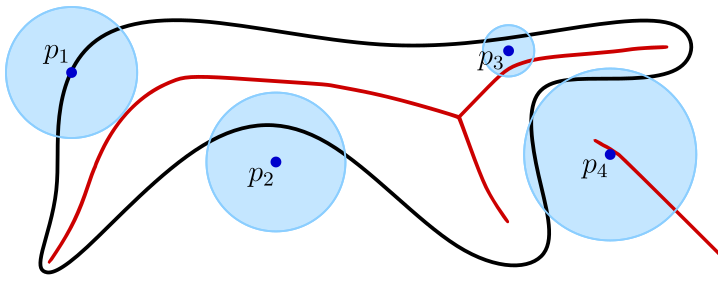}
    \caption{\revised{A planar curve (in black) and its medial axis (in red). Given some points (in dark blue), the disks centered at them (in light blue) either intersect the curve in a topological 1-disk, intersect the medial axis, or both.}}
    \label{fig:anytopodisk}
\end{figure}

In \Cref{fig:anytopodisk} we show an example of a planar curve intersected by Euclidean disks. At point $p_1$, the disk bounded by $\glfs{p_1}$ intersects the curve in a topological 1-disk, as imposed by \Cref{cor:curvetopodisk}. The other points depict the three possible cases imposed by \Cref{lem:anytopodisk2d} and its generalization - \Cref{lem:anytopodisk}: $p_2$ intersects the curve in a topological 1-disk; $p_3$ does the same, but it also contains part of the medial axis; $p_4$ intersects the curve in two disconnected components, and thus it must contain a part of the medial axis.
}

\paragraph*{Properties of the sampling.}

From \Cref{lem:anytopodisk} and \Cref{cor:curvetopodisk}, we can infer information on how the samples are distributed.
\begin{prp}\label{prp:nadj1reach}
    Let $S \subset \C$ be a sampling of the curve. For every point $p \in \C$, let $s_0, s_1 \in S$ be the samples such that the interval $I = (s_0, s_1)$ is the smallest open interval between samples that contains $p$. If there exists a point $q \in\C$ not belonging to the closure of $I$ that is closer to $p$ than both $s_0$ and $s_1$ are to $p$, then $d_{\M}(p, q) \geq \glfs{p}$.
\end{prp}
Note that the statement of \Cref{prp:nadj1reach} does not require that $p \notin S$. Indeed, if $p$ is a sample, the two samples that define the smallest interval containing $p$ are its adjacent samples. This is the reason we use the open interval containing $p$, and not the closed version. Given the proper constraints of $\rho$, we can provide additional guarantees.
\begin{cor}\label{cor:adjsample}
    Let $S \subset \C$ be a sampling of the curve, and $s_0, s_1 \in S$ be two adjacent samples defining an interval $I = [s_0, s_1] \subset \C$. If $S$ is a $\rho$-sampling with $\rho < 1$, then for every point $p \in I$, the closest sample to $p$ is either $s_0$ or $s_1$.
\end{cor}
\begin{prp}\label{prp:adj2reach}
    Let $S \subset \C$ be a $\rho$-sampling, with $\rho < 1$. For any two consecutive samples $s_0, s_1$, let $I = [s_0, s_1]$ be the interval between them. Then we have $d_{\M}(s_0, s_1) < 2 \greach{I}$.
\end{prp}

%% file: 03-method/02-sigdv/main.tex
\begin{figure}[t]
    \centering
    \includegraphics[width=0.33\columnwidth]{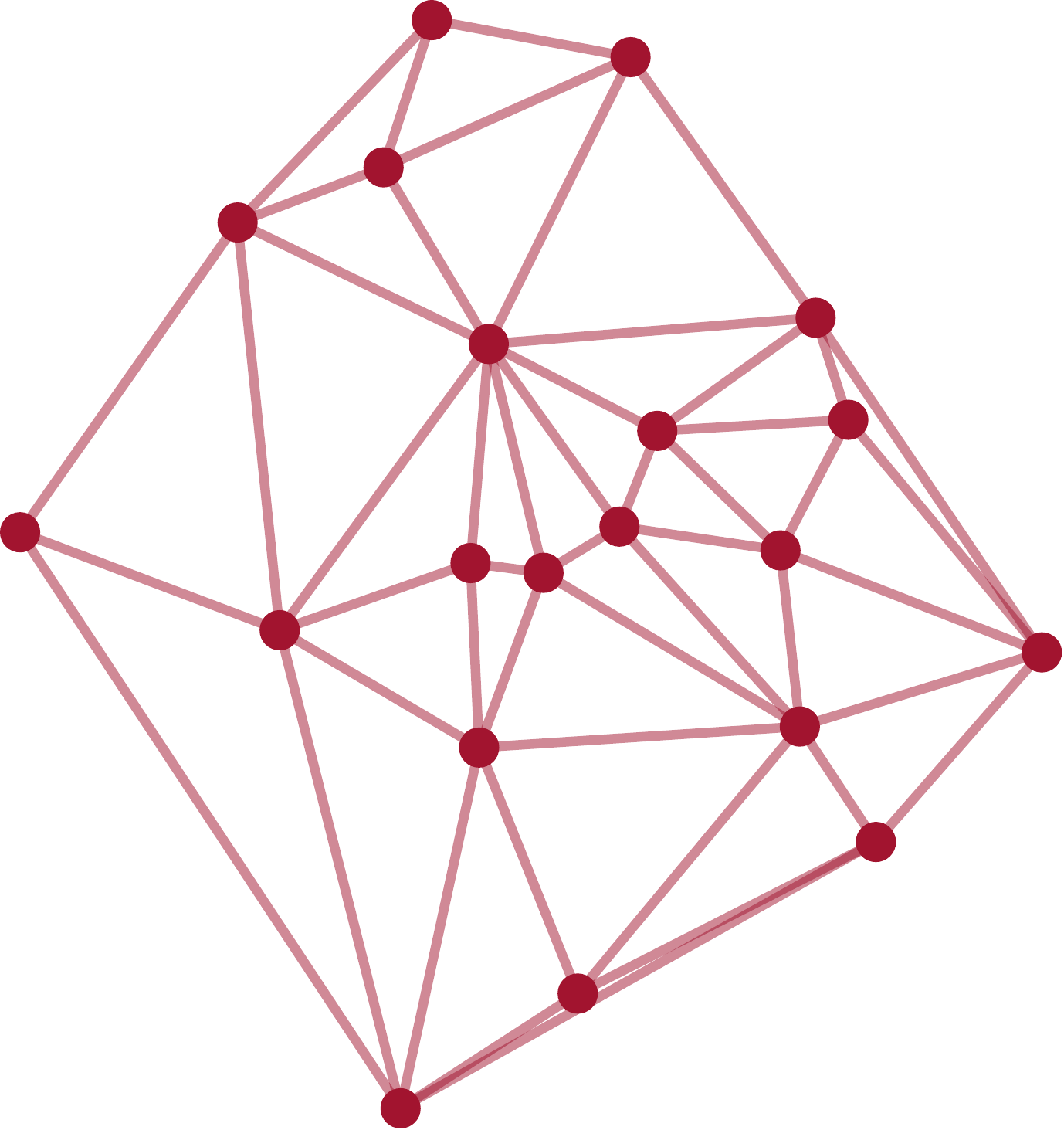}%
    \includegraphics[width=0.33\columnwidth]{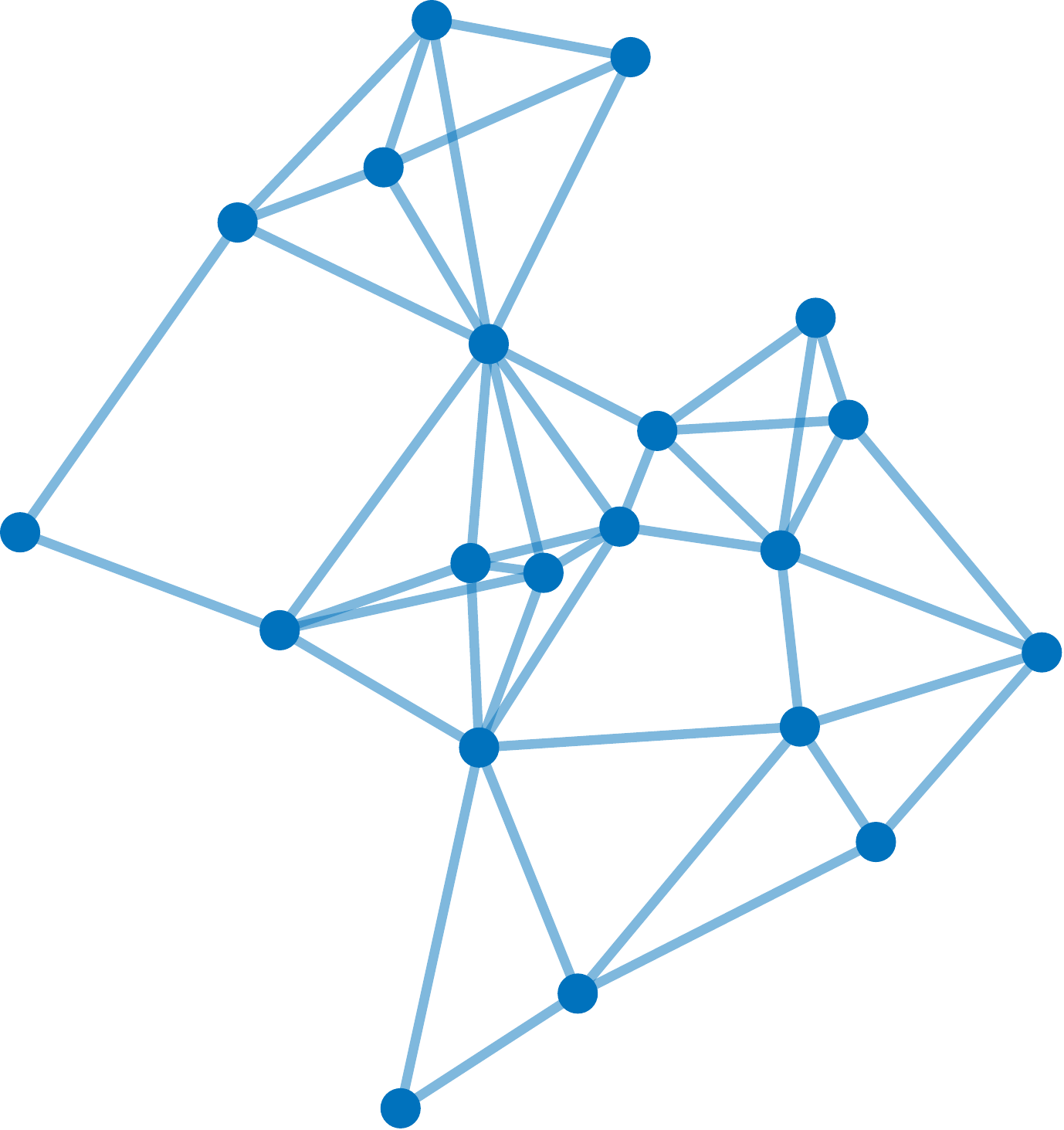}%
    \includegraphics[width=0.33\columnwidth]{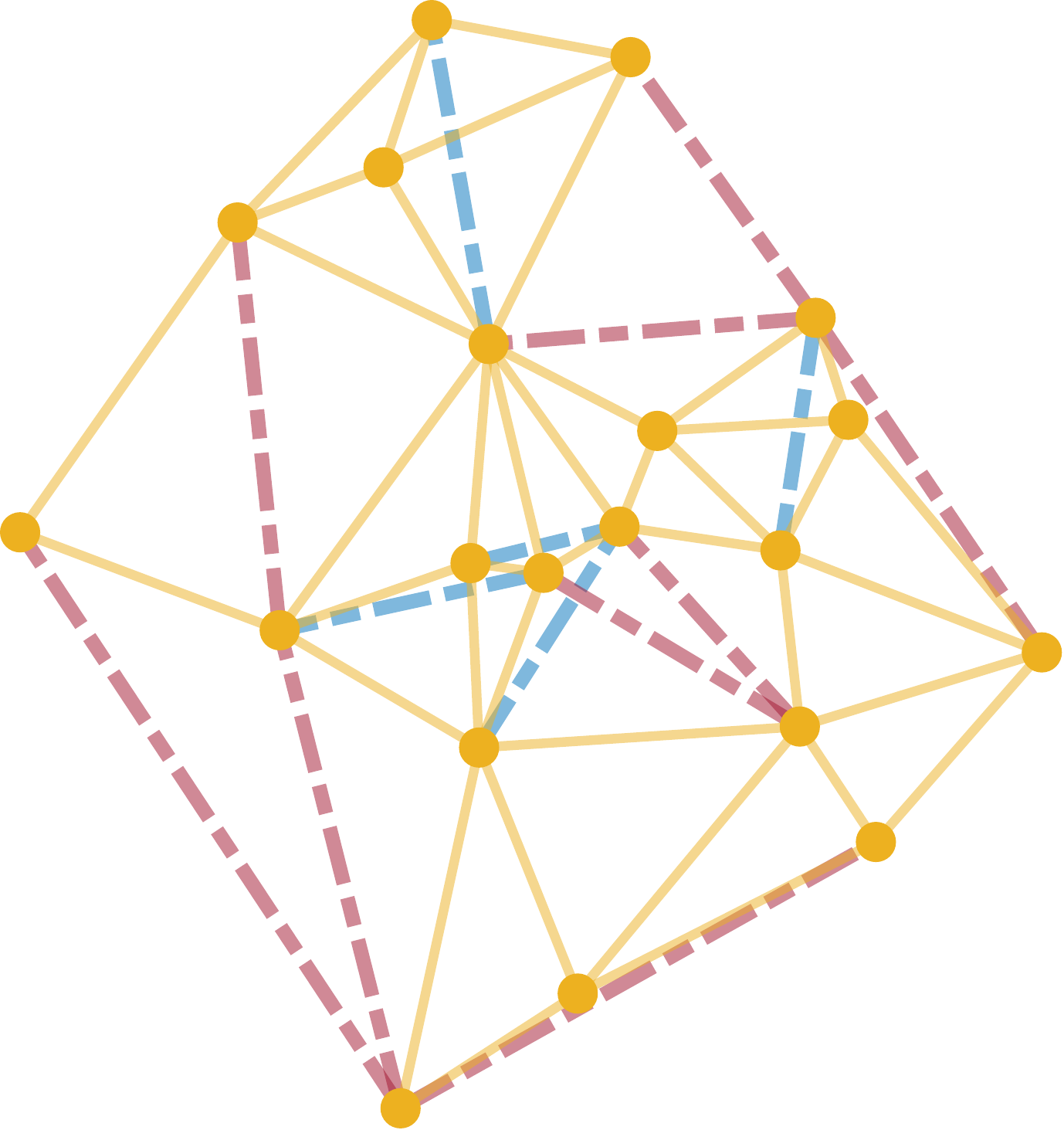}
    \caption{For a set of points in 2D, the intersection between Delaunay triangulation (red) and SIG (blue) creates the SIGDT (yellow). The edges removed from Delaunay triangulation and SIG are shown with dashed lines of the graphs' corresponding colors.}
    \label{fig:sigdt-example}
\end{figure}

In their work for 2D curve reconstruction, Marin \etal~\cite{marin:2022:sigdt} introduced the SIGDT proximity graph, obtained by intersecting the Delaunay triangulation with the Spheres-of-Influence graph.
\begin{dfn}[\hspace{1sp}\cite{toussaint:1988:sig}]\label{dfn:siggraph}
    Given a set of vertices $V = \{ v_1, \cdots, v_n \} \subset \M$, the \emph{Spheres-of-Influence (SIG) graph} of $V$ is a graph $G = (V, E)$ such that two nodes are connected if and only if they are closer than the sum of distances to their respective nearest neighbors. Namely
    \begin{equation}
        e = \{ v_i,\ v_j \} \in E
        \iff
        d_{\M}(v_i, v_j)
        \leq
        \delta_{NN}(v_i) + \delta_{NN}(v_j)
        \,,
    \end{equation}
    where $\delta_{NN}(v_i) = \min_{k \neq i} d_{\M}(v_i, v_k)$ is the distance of $v_i$ from its nearest neighbor.
\end{dfn}

The SIGDT is proved, in the planar setting, to contain the correct reconstruction of the curve under $\rho$-sampling conditions with $\rho < 1$ and non-uniformity ratio $u < 2$, and being a subgraph of the Delaunay triangulation, it is also a sparse graph. These conditions make it a good starting point for finding the reconstruction of the curve. We provide an example of the construction of the SIGDT in \Cref{fig:sigdt-example}.

While the Spheres-of-Influence graph is only defined in terms of distances, lifting the Delaunay triangulation to manifold domains is less intuitive. The lack of a coordinate system makes it difficult to generalize classical construction algorithms like the Bowyer-Watson method~\cite{bowyer:1981:delaunay,watson:1981:delaunay} to non-Euclidean metric spaces. However, we use the dual graph of the Delaunay triangulation -- the Voronoi partitioning \revised{of the vertex set}, which is defined only in terms of distances~\cite{aurenhammer:2013:voronoi}. This duality has already been exploited for lifting the Delaunay triangulation to surfaces and has proved to maintain many properties of the 2D triangulation, such as providing angle stability and containing the nearest neighbor graph~\cite{wang:2015:geodesicvoronoi,liu:2017:geodesicdelaunay}. Furthermore, it can be shown that computing the Voronoi partitioning of discrete manifolds is an efficient operation that can be achieved in time $\BigO{|V| \log|V| \log k}$, being $|V|$ the number of vertices and $k$ the number of samples~\cite{peyre:2006:frontpropagation,maggioli:2024:rematching}.

\begin{figure}[t]
    \centering
    \includegraphics[width=0.33\columnwidth]{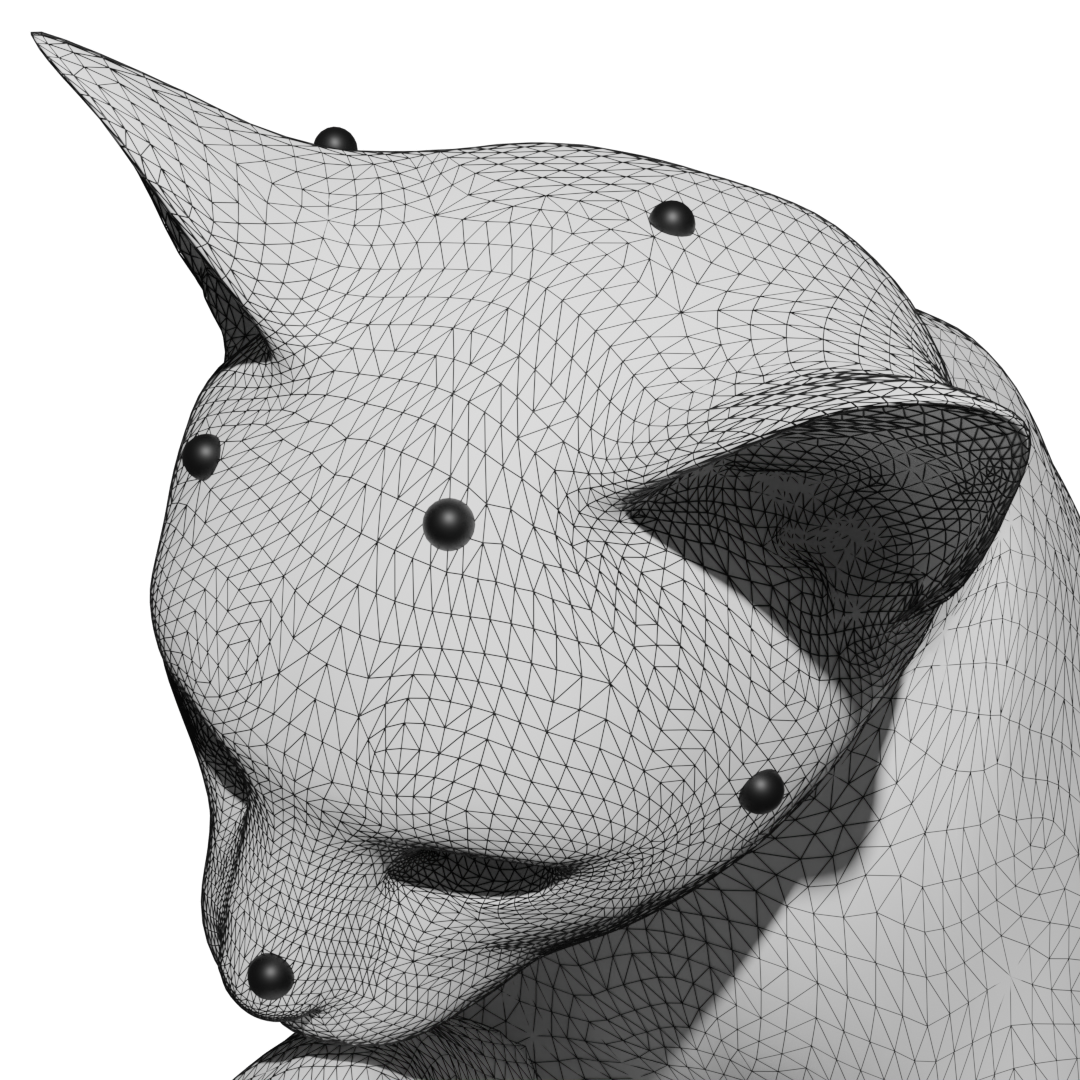}%
    \includegraphics[width=0.33\columnwidth]{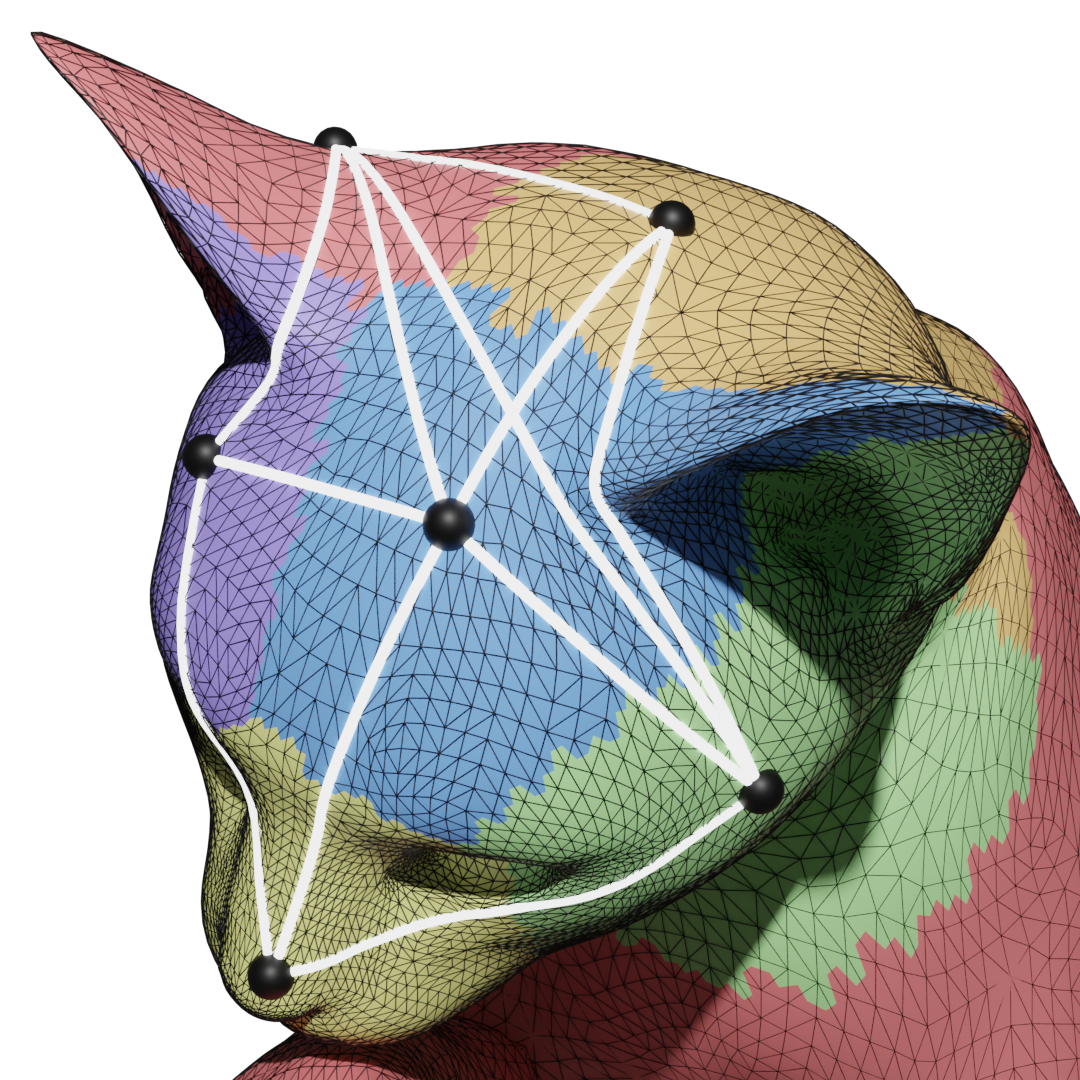}%
    \includegraphics[width=0.33\columnwidth]{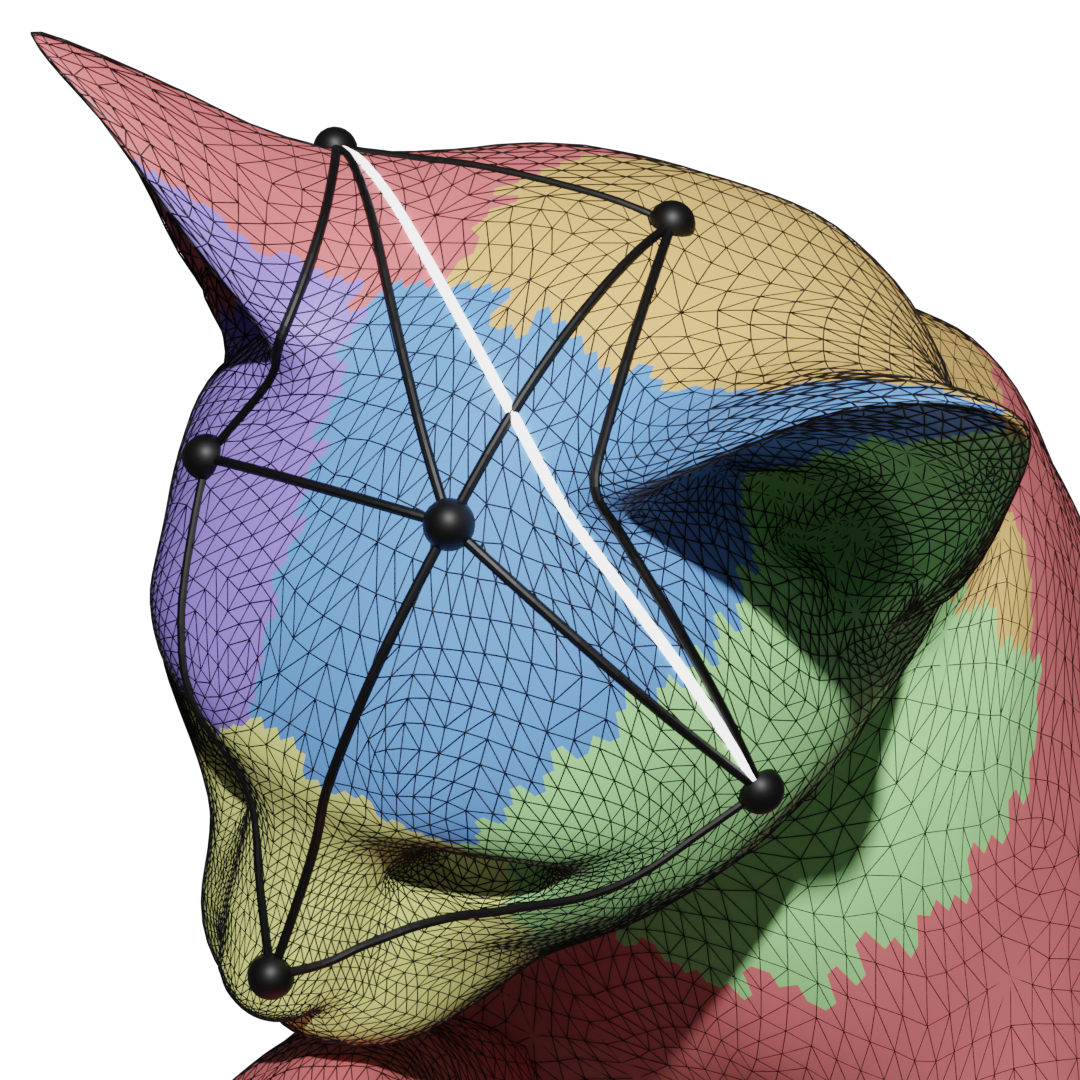}
    \caption{Left: a triangular mesh with some surface samples. Center: the geodesic Voronoi partitioning induced by the samples on the mesh, and its dual graph (white) obtained by locating the shortest geodesic paths between samples in adjacent partitions. Right: the SIGDV (black) obtained by removing the edges not in the SIG.}
    \label{fig:dual-voronoi-example}
\end{figure}

Differently from the Euclidean space, the dual of a Voronoi decomposition does not always guarantee a valid triangulation of the point set, as there are some additional constraints that the samples must satisfy~\cite{amenta:2000:closedball,dyer:2007:voronoi}. If these are not satisfied, the dual still forms a sparse graph that encodes the proximity of samples well. In order to obtain the dual of the Voronoi, we connect samples that generate adjacent partitions, as shown in \Cref{fig:dual-voronoi-example}.
By removing all the edges from the Voronoi dual that do not also satisfy \Cref{dfn:siggraph}, we obtain the intersection between the Spheres-of-Influence graph and the dual Voronoi graph, which we denote as \emph{SIGDV}, and which provides a generalization for the SIGDT graph.

The following statements about our graph hold. Proofs are provided in \Cref{sec:proofs}.
\begin{lem}\label{lem:dvedge}
    Let $s_0, s_1 \in \C$ be consecutive samples from a $\rho$-sampling $S \in C$, with $\rho < 1$. The edge $e(s_0, s_1)$ belongs to the dual Voronoi graph of $S$.
\end{lem}

\begin{lem}\label{lem:sigedge}
    Let $s_0, s_1, s_2, s_3 \in \C$ be consecutive samples from a $\rho,u$-sampling $S \in C$, with $\rho < 1$ and $u < 2$. The edge $e(s_1, s_2)$ belongs to the Spheres-of-Influence graph of $S$.
\end{lem}

Finally, we prove (in \Cref{sec:proofs}) that SIGDV contains the correct reconstruction under analogous sampling conditions as the 2D case:
\begin{thm}\label{thm:sigdvedge}
    Let $\M$ be a $d$-dimensional Riemannian manifold, possibly with boundary $\partial\M$, and equipped with a geodesic distance $d_{\M} : \M\times\M \to  \Reals$. Let $\C \subset \M$ be a curve on the manifold $\M$, and $S = \{ s_1, \cdots, s_k \} \subset \C$ be a $\rho,u$-sampling of $\C$, with $\rho < 1$ and $u < 2$. The edge connecting any pair of consecutive samples is part of the SIGDV of $S$.

\end{thm}

\revised{%
\begin{figure}[t]
    \centering
    \includegraphics[width=0.33\columnwidth]{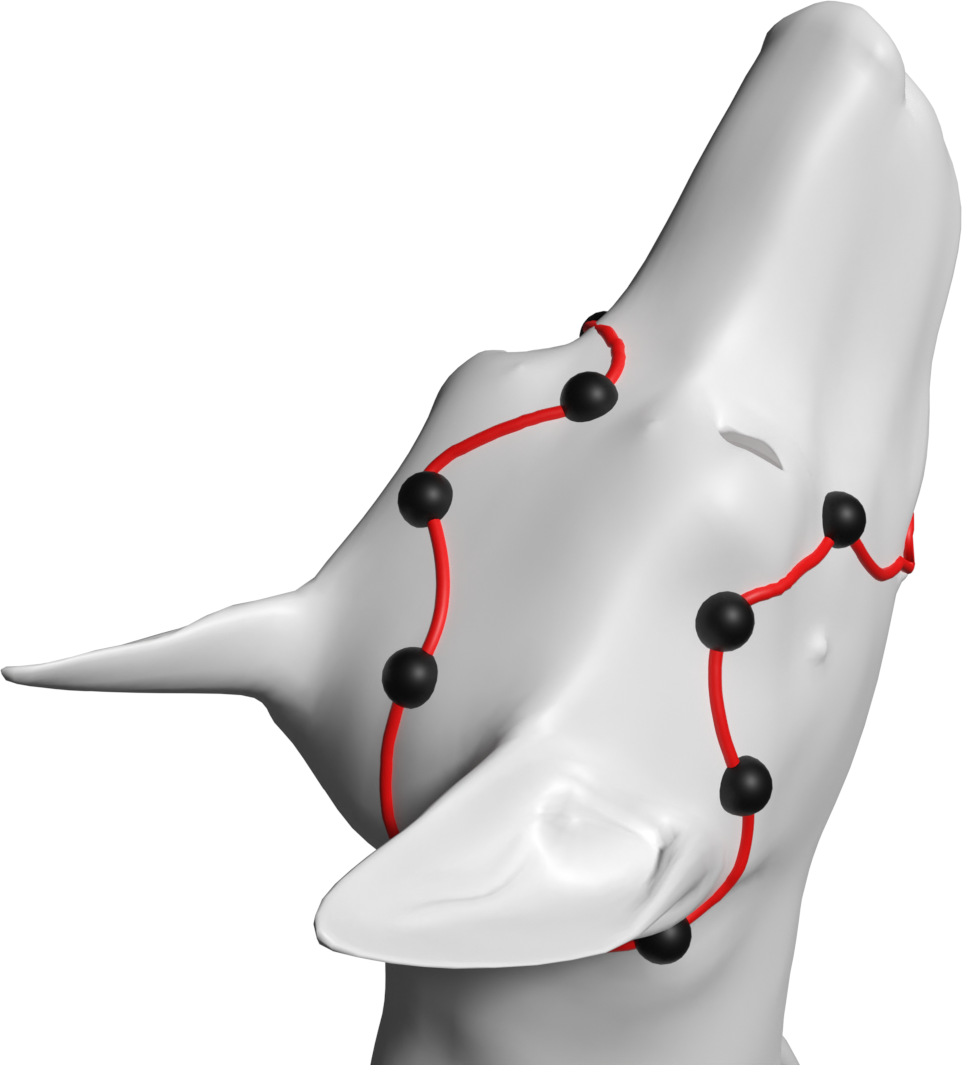}%
    \includegraphics[width=0.33\columnwidth]{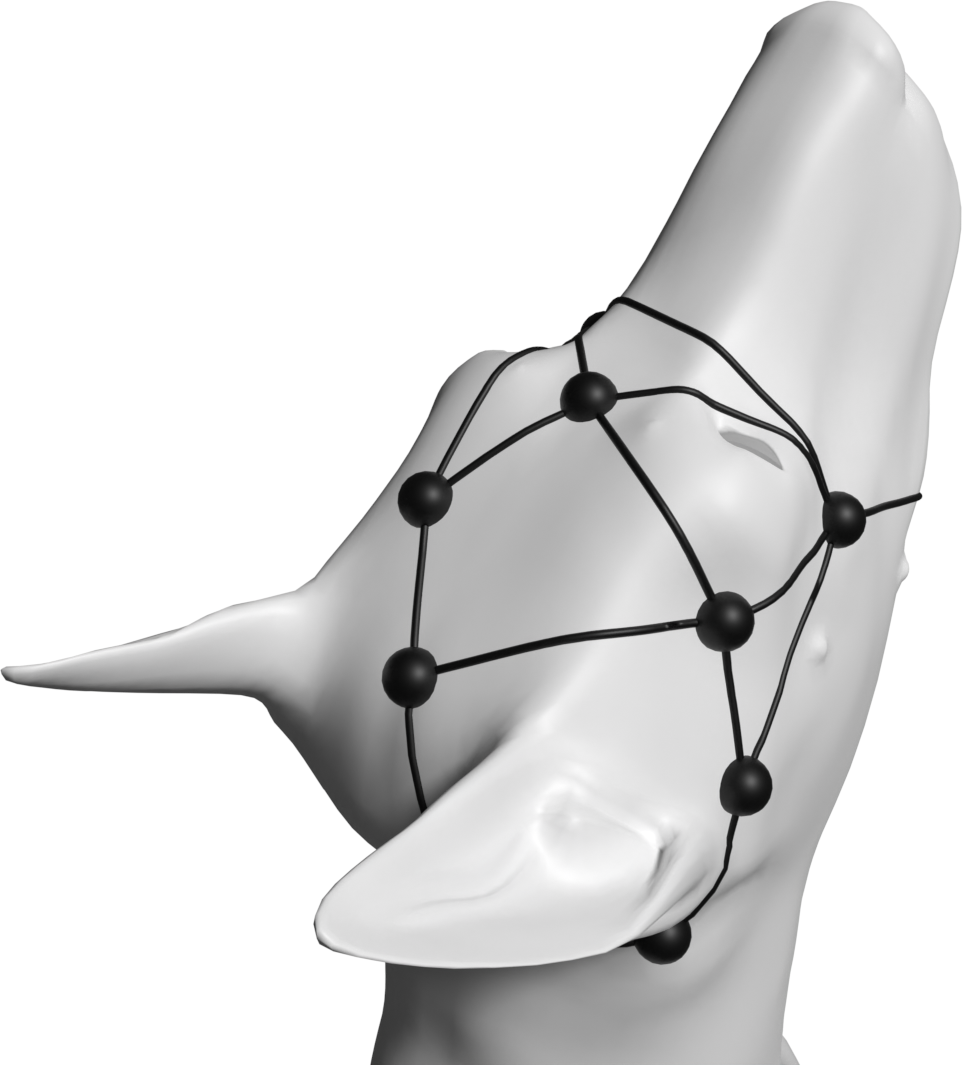}%
    \includegraphics[width=0.33\columnwidth]{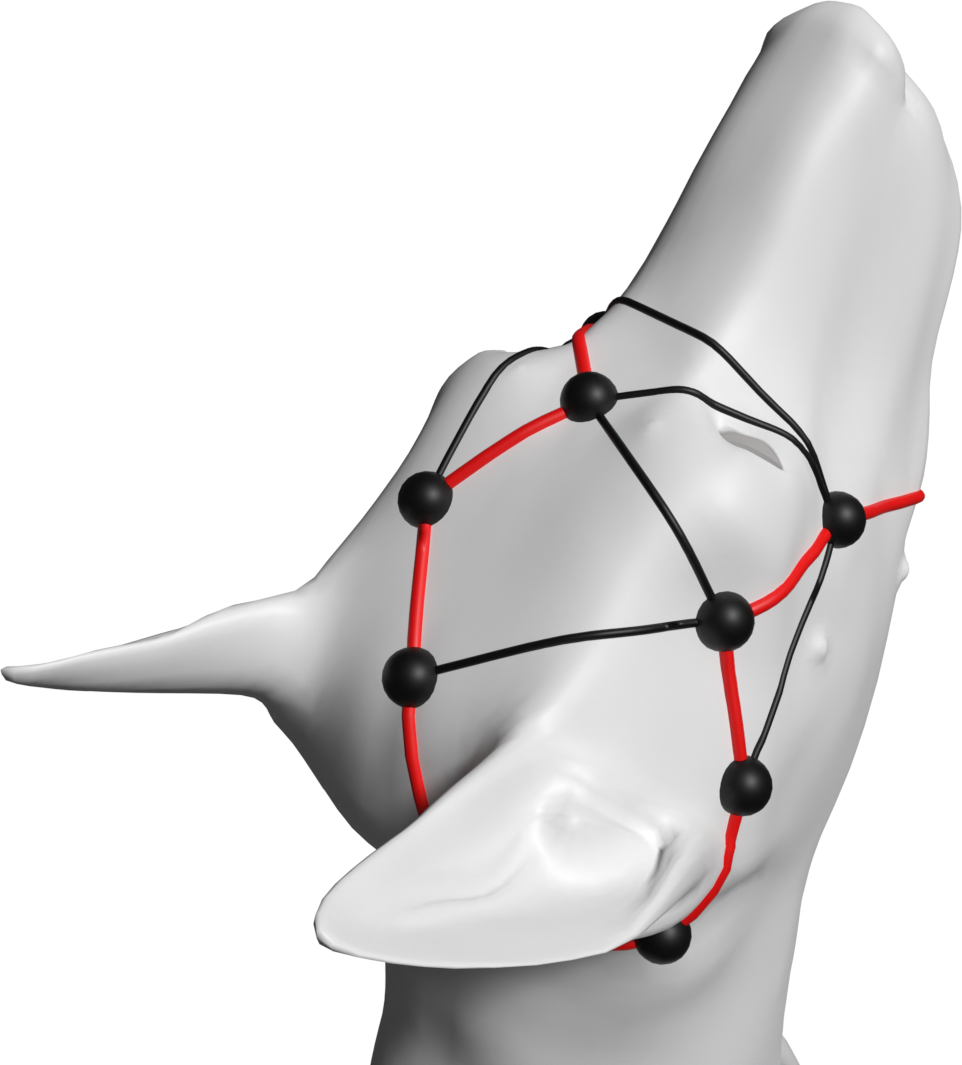}
    \caption{\revised{Left: a smooth closed curve (red) sampled at some vertices (black). Middle: the vertices are connected using SIGDV (black), including the edges between consecutive samples. Right: the shortest Hamiltonian path (red) inside SIGDV.}}
    \label{fig:sigdvedge}
\end{figure}

In \Cref{fig:sigdvedge} we illustrate an example application of \Cref{thm:sigdvedge} to a curve on a 2-dimensional surface. The smooth curve wrapping around the head of the dog shape is sampled with a set of points, which are then connected with SIGDV. If two samples are adjacent on the curve, they share an edge inside SIGDV.

}

%% file: 03-method/03-tsp/main.tex
While \Cref{thm:sigdvedge} guarantees that SIGDV contains the correct reconstruction, it may contain additional edges. This problem was also addressed by Marin \etal~\cite{marin:2022:sigdt} for their SIGDT planar curve reconstruction algorithm. In their pipeline, the authors propose to apply a series of inflating and sculpting operations until they obtain a manifold curve. In our case, this is not applicable, as in non-Euclidean spaces, such as manifolds, it is not always possible to define the notions of ``inside'' and ``outside'' of a curve.

To overcome this issue, we note that the edges of the correct reconstruction should create a cycle such that each sample is visited only once. Various ways of extracting these edges exist in the literature, such as the identification of a Hamiltonian cycle~\cite{iwama:2007:backtrackingtsp,bjorklund:2014:montecarlohamiltonian} or the solution to a Traveling Salesman Problem (TSP). \revised{While we are not guaranteed that the TSP solution encodes the ground truth ordering in general, it has been shown that outputting a shortest path adheres to the Gestalt principles and provides a good heuristic in the planar case~\cite{althaus:2000:tspcurves,ohrhallinger:2013:connect2d,ohrhallinger:2021:benchmark}.
In our experiments, this heuristics proved to generalize well on manifold domains, as shown in the example from \Cref{fig:sigdvedge}.
}

The Traveling Salesman Problem is a fundamental problem in computer science, and it is at the very core of many optimization problems. Given a set of points, TSP requires finding the shortest path that connects all the points, thus solving the problem of curve reconstruction as well. The problem can be formulated in different spaces and with different connectivity constraints, but all of them have been shown to be NP-hard~\cite{karp:1975:npcproblems}. However, the importance of the TSP led many researchers to deploy approximated solutions, especially relying on a nearest-neighbor approach~\cite{johnson:1997:nntsp,ray:2007:fntsp} -- even though these solutions tend to have a large approximation factor. Other techniques~\cite{dorigo:1996:ant,sathya:2015:tspsurvey} proved to be much more effective, while paying a small price on time efficiency. 

Most of the research related to the TSP assumes complete graphs (\ie, each node is connected to every other node). While we can accept solutions containing edges not belonging to SIGDV, we know that this graph contains the correct reconstruction under some sampling conditions, and hence we want to bias the algorithm to use SIGDV edges, while at the same time guaranteeing an efficient solution.

We start from a solution that exploits the Minimum Spanning Tree of a graph to produce a 2-optimal approximation to the problem (\ie, a solution that is at most two times as expensive as the optimal one)~\cite{cormen:2022:algorithms}.
Our method is summarized in \Cref{algo:sigdv-tsp}, where the input graph $G$ is the SIGDV graph and the matrix $\mymat{D}$ contains in the entry $\mymat{D}_{ij}$ the geodesic distance $d_{\M}(i, j)$ between the $i$-th and the $j$-th samples.

\begin{algorithm}[t]
    \caption{Algorithm for solving the TSP biased towards SIGDV edges.}\label{algo:sigdv-tsp}
    \begin{algorithmic}[1]
        \Procedure{TSPSolver}{$G = (V, E)$, $\mymat{D} \in \Realsn{|V|\times|V|}$}
            \State $T \gets$ minimum spanning tree of $G$ w.r.t. $\mymat{D}$
            \State $P \gets$ ordering of $V$ following a pre-order DFS visit of $T$
            \Repeat
                \State Find $(i, j), (\ell, h) \in P$ s.t. $\mymat{D}_{i j} + \mymat{D}_{\ell h} > \mymat{D}_{i \ell} + \mymat{D}_{jh}$
                \State $P \gets P \setminus \{ (i, j), (\ell, h) \}$
                \State $P \gets P \cup \{ (i, \ell), (j, h) \}$
            \Until{$P$ is unchanged}
            \State \Return $P$
        \EndProcedure
    \end{algorithmic}
\end{algorithm}

We first compute the minimum spanning tree $T$ of the graph $G$ and via a pre-ordered Depth-First Search (DFS) visit of $T$ we find an ordering of the nodes in $V$ that we use to build an initial cycle $P$ (Lines 1-2). Since the metric $d_{\M}$ respect the triangular inequality, the cycle $P$ is guaranteed to be a 2-optimal approximation of the TSP~\cite{cormen:2022:algorithms}. However, $P$ does not offer any guarantee for the cycle to be a local minimum of the problem. Thus, we post-process the ordering by searching for any pair of edges $(i, j), (\ell, h)$ such that if we swap them into $(i, \ell), (j, h)$ we obtain a shorter cycle, and we iterate this process until no new pair of edges is found (Lines 4-8). 
\revised{This edge-swap procedure was originally proposed by Croes~\cite{croes:1958:tsp} for solving the TSP with an approximation factor of $\BigO{\nicefrac{\log n}{\log \log n}}$ on Euclidean instances (being $n$ the number of vertices)~\cite{brodowsky:2023:approximation}.}

\paragraph*{Relaxed sampling conditions.}
The sampling conditions imposed by \Cref{thm:sigdvedge} allow for a sparser sampling compared to the state-of-the-art. 
However, when these conditions are not met, we do not have the guarantee that SIGDV contains the correct reconstruction or any other Hamiltonian cycle. 
Intersecting the dual Voronoi graph with the SIG could even produce a graph with multiple disconnected components. In some applications (see \Cref{fig:teaser}) \revised{the input could be known to include samples from distinct closed loops and}
the clustering can be used to identify multiple curves. \revised{If the user desires to reconstruct} only a single curve, we alter the graph to produce a single connected component. First, we build the complete graph $\mathfrak{G} = (\mathfrak{V}, \mathfrak{E})$ of the connected components, where each node $N_i \in \mathfrak{V}$ represents a single component of SIGDV, and the length of the edge $(N_i, N_j) \in \mathfrak{E}$ is given by $\min_{u \in N_i, v \in N_j} d_{\M}(u, v)$. 
Then, we compute the minimum spanning tree $\mathfrak{T}$ of this graph. For every edge $(N_i, N_j) \in \mathfrak{T}$, we add to SIGDV the edge $(u^*, v^*) = \argmin_{u \in N_i, v \in N_j} d_{\M}(u, v)$.
This pre-processing guarantees that we obtain a single connected component, while, at the same time, adding non-SIGDV edges with minimum total edge length.
The choice between single and multiple curve reconstruction is left to the user for full flexibility.

%% file: 04-results/main.tex
We investigate several diverse applications of our method. We first compare it to the approach proposed by Shah \etal~\cite{shah:2013:curve} (MST) for motion tracking applications, where we show that we are able to reconstruct complex paths with far fewer samples. We show how our method can be applied in processing archaeological data, and how it can improve contour matching results, and we discuss the applicability of our solution to isoline extraction from sparse samples for scientific visualization.

Our method has been implemented in C++, using Eigen~\cite{eigen} and Geometry Central~\cite{geometrycentral}. \revised{The implementation is available at \href{https://github.com/filthynobleman/curves-surf}{https://github.com/filthynobleman/curves-surf}.} For computing Voronoi diagrams in high-dimensional spaces we use Qhull~\cite{barber:1996:quickhull}, and for computing the geodesic paths used in our visualizations we use the Flip Geodesics algorithm~\cite{sharp:2020:flipout}.

\subsection{Motion tracking}\label{sec:motion}

\input{04-results/04-motion/main}

\subsection{Virtual cultural heritage}\label{sec:qualitative}
\input{04-results/02-archaeo/main}

\subsection{Contour matching}\label{sec:contour}

\input{04-results/01-contour/main}

\subsection{Sparse data visualization}\label{sec:isolines}

\input{04-results/05-isotemp/main}


%% file: 04-results/04-motion/main.tex
In their work about curve reconstruction on Riemannian manifolds, Shah \etal~\cite{shah:2013:curve} propose, as an application, the reconstruction of curves in the space of rigid motions $\se{3}$ for tracking the motion of an object from position and rotation samples. To apply our method to this task, we first need to define how to compute a Voronoi diagram in $\se{3}$.

As discussed by Gonz\'{a}lez-L\'{o}pez \etal~\cite{gonzalez:1995:voronoise3}, the group $\se{3}$ can be expressed as the product $\so{3}\times\Realsn{3}$, $\so{3}$ being the group of rotations in 3D space. The authors propose a method for computing the Voronoi decomposition in $\so{3}$ relying on the following result.
\begin{prp}[\hspace{1sp}\cite{gonzalez:1995:voronoise3}]
    Given a set of rotations $R = \{ r_i \}_{i = 1}^{k}$ in $\so{3}$ and a set of rotation matrices $R' = \{ \mymat{R}_i \}_{i = 1}^{k}$ in $\Realsn{3\times3}$ such that $\mymat{R}_i$ represents the rotation of $r_i$, the Voronoi decomposition $\mathrm{Vor}(R, d_{\so{3}})$ of $\so{3}$ w.r.t. the rotations $R$ can be obtained as
    \begin{equation}
        \mathrm{Vor}(R, d_{\so{3}})
        =
        \mathrm{Vor}(R', d_{\Realsn{9}})
        \cap
        \so{3}\,.
    \end{equation}
\end{prp}

We improve this result for an easier computation of the Voronoi diagram in the group $\so{3}$. The proof is provided in \Cref{sec:proofs}.
\begin{prp}\label{prp:so3r4}
    Given a set of rotations $R = \{ r_i \}_{i = 1}^{k}$ in $\so{3}$ and a set of quaternions $Q = \{ q_i \}_{i = 1}^{k}$ in $\mathbb{H}$ such that $q_i$ represents the rotation $r_i$, then the Voronoi decomposition $\mathrm{Vor}(R, d_{\so{3}})$ of $\so{3}$ w.r.t. the rotations $R$ can be obtained as
    \begin{equation}
        \mathrm{Vor}(R, d_{\so{3}})
        =
        \mathrm{Vor}(Q, d_{\Realsn{4}})
        \cap
        \so{3}\,.
    \end{equation}
\end{prp}

As proven by Gonz\'{a}lez-L\'{o}pez \etal~\cite{gonzalez:1995:voronoise3}, this result does not generalize to $\se{3}$, and we must only rely on the Voronoi decomposition of the embedding space. However, by reducing the dimensionality of the embedding space of $\so{3}$ from $\Realsn{9}$ to $\Realsn{4}$, we also reduce the dimensionality of the embedding space of $\se{3}$ from $\Realsn{12}$ to $\Realsn{7}$. Thus, we can use the Voronoi decomposition in $\Realsn{7}$ as a reasonable approximation of the Voronoi decomposition in $\se{3}$.

\begin{figure}[t]
    \centering
    \includegraphics[width=\columnwidth]{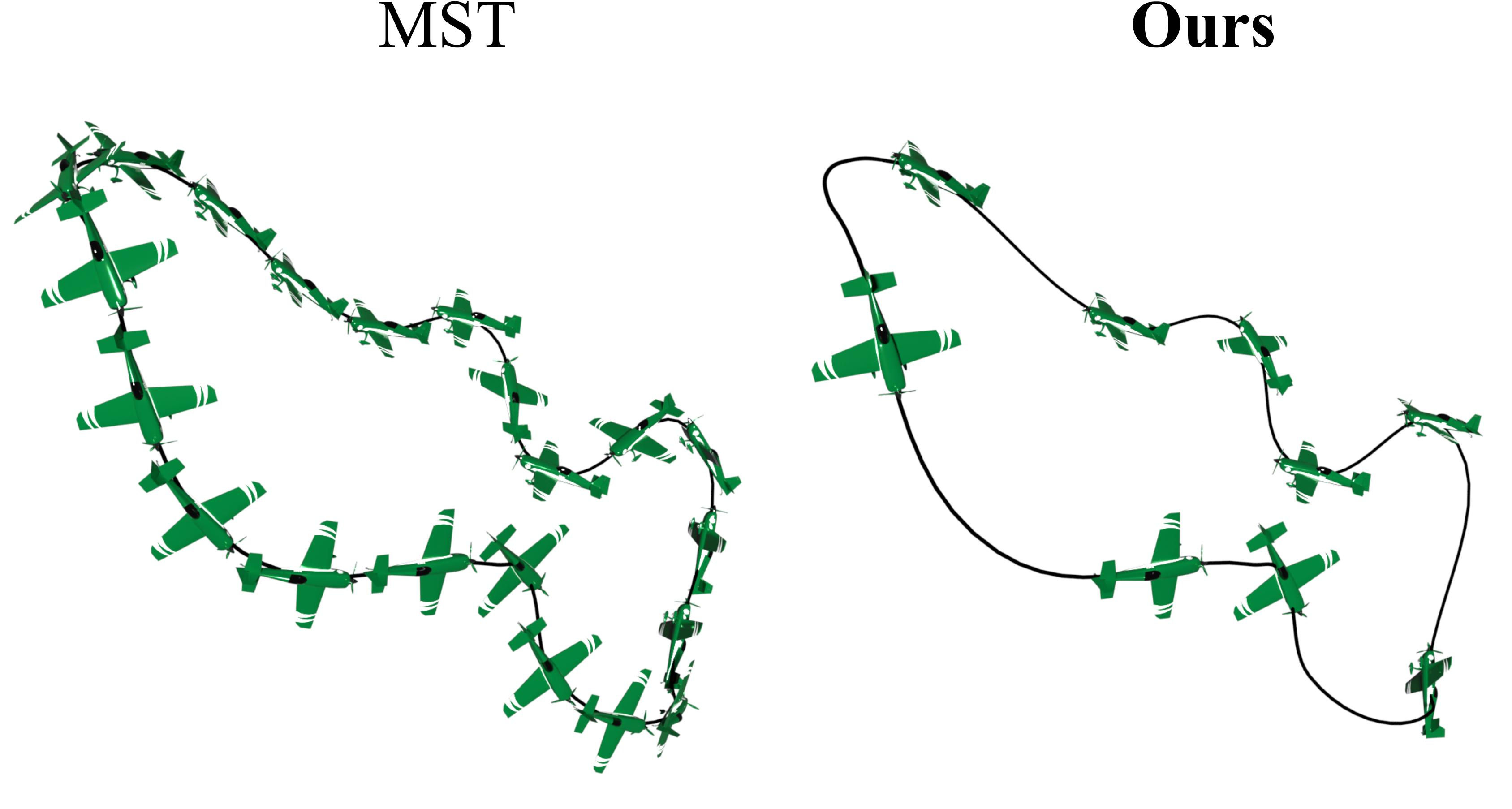}
    \caption{Minimal number of samples for reconstructing the motion of an airplane along a path using MST - left (21), and our approach - right (9). \revised{The curve shows the ground truth path of the airplane.} The \href{https://skfb.ly/oCJRy}{airplane model} has been created by Kemal Çolak and distributed by Sketchfab under the license CC BY 4.0.}
    \label{fig:motion-tracking-airplane}
\end{figure}

In \Cref{fig:motion-tracking-airplane} we present an example of motion reconstructed using MST and our solution, and we show the minimum number of samples required by each method to \revised{correctly reconstruct the motion sequence} -- 21 for MST and 9 for our method. While our method can deal with sparse non-uniform sampling and still correctly recover the ordering of the samples, it is evident here that MST requires a more dense and uniform sampling scheme. Indeed, the path presents a challenging shape, as the two close turns in the bottom half are very close in space, and the adjacent samples on the curve are very different in terms of rotations. Such curves can be adversarial for MST, as the distance in $\se{3}$ between non-adjacent samples is small, preventing their method from finding a closed curve.

%% file: 04-results/02-archaeo/main.tex
\begin{figure}[t]
    \centering
    \includegraphics[width=\columnwidth]{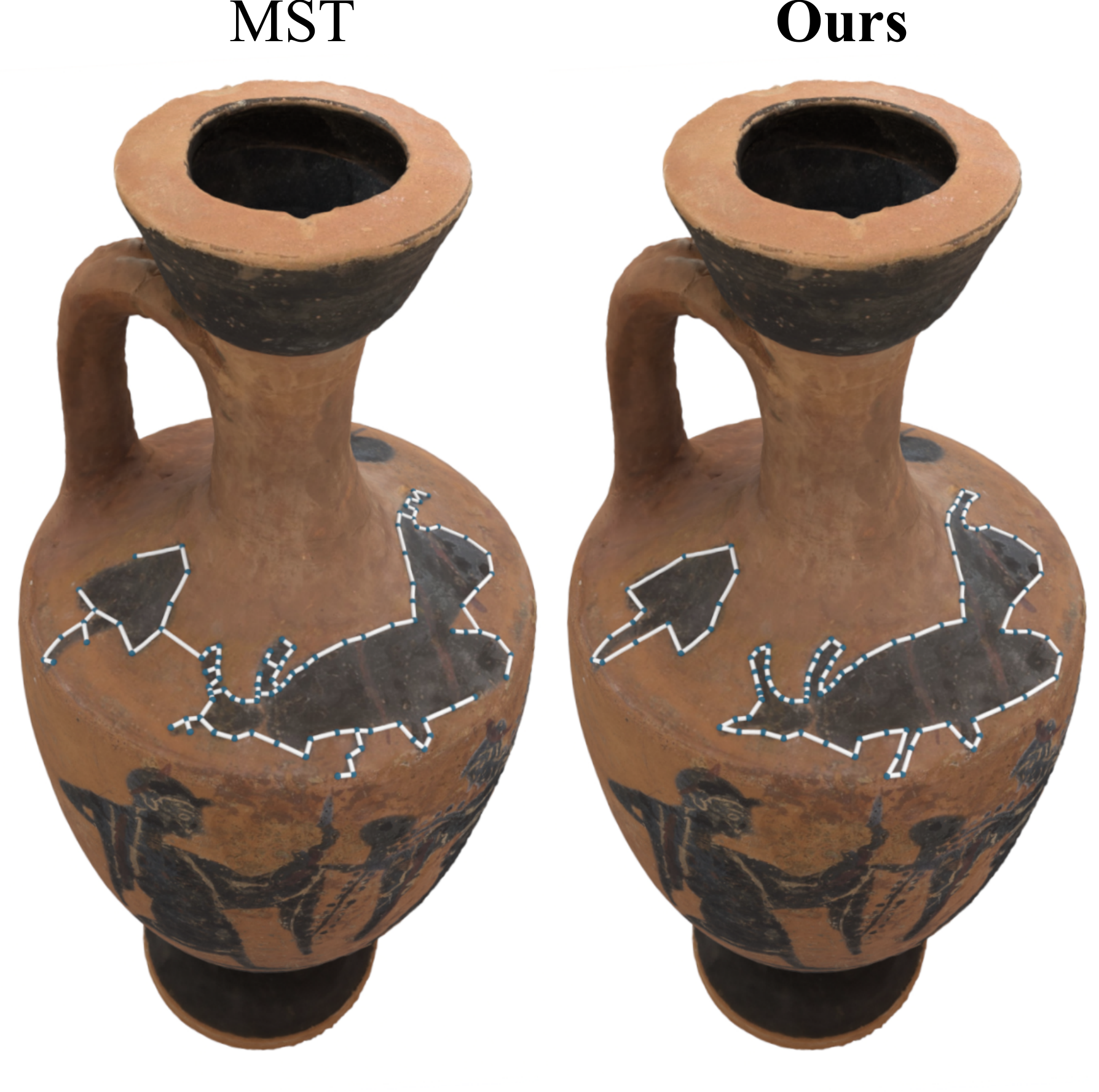}
    \caption{Reconstruction of complex curves (white) on the surface of a clay vase from a set of manually chosen samples (blue). The MST algorithm (left) fails to reconstruct meaningful contours, while ours (right) correctly identifies the two shapes. The \href{https://skfb.ly/XF8F}{vase model} has been created by Laura Shea and distributed by Sketchfab under the license CC BY-NC-SA 4.0 DEED.
    }
    \label{fig:archeo}
\end{figure}

Scientific visualization and 3D shape analysis are already prevalent in archaeological applications as instruments for handling the always-increasing amount of data available to cultural heritage research~\cite{tal:2014:3darchaeo}.

Successful applications of computational geometry to the analysis of archaeological data include the identification of demarcating curves on low reliefs, statues, and other various kinds of cultural heritage artifacts~\cite{gilboa:2013:curvearcheo,torrente:2016:featurecurves}, curve matching tasks in 2D and 3D for solving the problem of fragment reconstruction~\cite{mcbride:2003:archaeologicalcurvematching} and pattern extraction of culturally significant patterns~\cite{yao:2020:feature}.

The problem of reconstructing curves on surfaces is closely related to these applications, as there can be cases where missing information from lost pieces could be inferred by knowing the structure of the underlying artifact (\eg partial designs on vases that could be reconstructed from still existing pieces). In \Cref{fig:archeo} we show that our method can faithfully reconstruct complex shapes on the surface of a vase, precisely contouring the black drawings from a sparse sampling. We also compare our result against the MST algorithm, which is unable to find a closed shape due to the presence of multiple connected components and the large distance among some of the samples.

%% file: 04-results/01-contour/main.tex

Contour matching is a novel task in computer graphics and vision applications that has recently achieved high attention~\cite{lahner:2016:elastic2d3d,roetzer:2023:conjugategraph}. The problem this field is concerned with is how to meaningfully relate a surface to the curve defining its contour, even under non-rigid deformation of the surface~\cite{windheuser:2011:elastic2d3d}.

While the existing approaches reliably compute non-rigid correspondences between shapes and their contour, the task is still new and even state-of-the-art methods do not always provide high-quality guarantees, producing self-intersecting and locally degenerate curves. We show that our curve reconstruction technique can be exploited as a refinement step to improve the contour quality. 

For our evaluation, we use the method proposed by L\"{a}hner \etal~\cite{lahner:2016:elastic2d3d} (\textsc{Elastic2D3D}), selecting some instances from the dataset that the authors derive from the FAUST shape collection~\cite{bogo:2014:faust} where the method produces many degeneracies. From the curve obtained with \textsc{Elastic2D3D}, we randomly extract 3\% of the vertices, and we run our algorithm for reconstructing the contour over the target shape. 

\begin{figure}[t]
    \centering
    \includegraphics[width=\columnwidth]{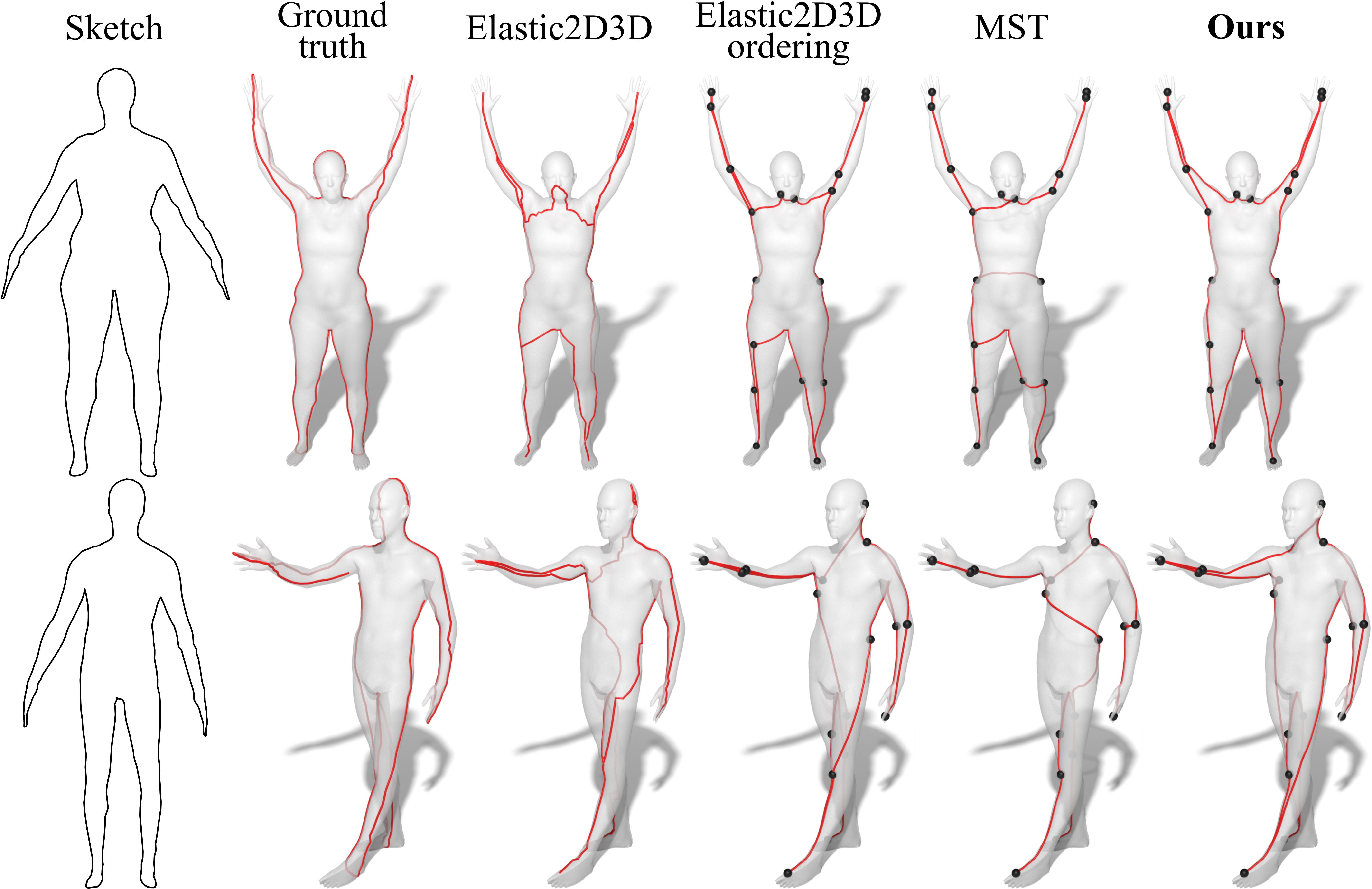}
    \caption{Examples of how our curve reconstruction method is applied to refine a contour. Given a contour obtained with \textsc{Elastic2D3D}~\cite{lahner:2016:elastic2d3d}, we sample it sparsely and reconnect the samples using the ordering of the contour, MST, and our algorithm.}
    \label{fig:contour-samples}
\end{figure}

Even if \textsc{Elastic2D3D} may produce self-intersections and locally degenerate curves, it captures the overall shape of the contour. Thus, a sparse sampling of the curve should give a good sparse contour. As shown in the examples from \Cref{fig:contour-samples}, with our algorithm we can smooth the results of \textsc{Elastic2D3D}, resolving the self-intersections and producing non-degenerate curves. For reference, we also connect the samples with their original ordering, proving that our method's reordering of the samples improves the overall result. 

MST cannot produce a valid closed curve in any of the tested cases, as the conditions imposed by the algorithm require the sampling to be denser and more uniform for the minimum spanning tree to result in a chain. This is primarily due to the presence of degeneracies in the curve produced by \textsc{Elastic2D3D} in the form of jagged edges. Hence, two points sampled from a degenerate section of the curve would very likely result in a branch on the minimum spanning tree. We validate this argument by running the MST algorithm on different sampling densities spanning from 1\% to 100\% of the curve vertices produced by \textsc{Elastic2D3D}, and the minimum spanning tree never results in a chain.

%% file: 04-results/05-isotemp/main.tex
Having shown that our method can improve contours already computed by reordering a sparse set of their samples, we now move onto another definition of contours, namely their existence as isolines (\ie, lines connecting data points with the same value). We show how our method can be used to visualize large data by only using a sparse subset of it.

Analyzing large datasets and being able to extract meaningful visual information is a challenge for various scientific visualization fields. We choose to focus on meteorological data in this application, and considering the increase in computational power and available tools, the information we can obtain has increased exponentially. Methods to quickly process massive amounts of data are required, to allow scientists to interpret and extract the most relevant information. One method to deal with very dense data is to choose a subset, that acts as an overview, and then use this global view of data to find and focus on a specific area of interest at a higher resolution~\cite{kim:2019:sampling,mahmud:2020:surveysampling}. \revised{In the case of prohibitively large datasets, such as hourly records of multiple weather parameters, processing solely a subset of the data might be the only way to manipulate information at such a scale.} This facilitates a quicker understanding of the data at different levels while using less data and hence, less computational power. 

In the field of visualizing meteorological data, an important problem arises from the numerical issues of projecting information that exists on the Earth's surface onto a plane, where at least one of the following properties has to be sacrificed: distances, angles, or areas. We bypass this challenge by working with samples directly on surfaces and allowing for a three-dimensional interaction and visualization of data. A common way of visualizing various properties of meteorological data is presented by contour lines (or isolines) where data points with equal or similar (up to a threshold) values are connected to represent all the possible points that have similar data~\cite{rautenhaus:2017:vizmeteo}.

Meteorological datasets, such as the ones provided by the European Centre for Medium-Range Weather Forecasts~\cite{ecmwf}, contain massive amounts of data for every recorded timestep, at a resolution of 0.25 degrees in both latitude and longitude, with up to 90 parameters such as temperature, humidity, pressure. Analyzing the entirety of the data at a global level is a tedious and complex task, and we propose to combine the deformation-free surface representation with a sparse sampling of the dataset to reconstruct contours for an easier understanding of global information.

\begin{figure}[t]
    \centering
    \includegraphics[width=\columnwidth]{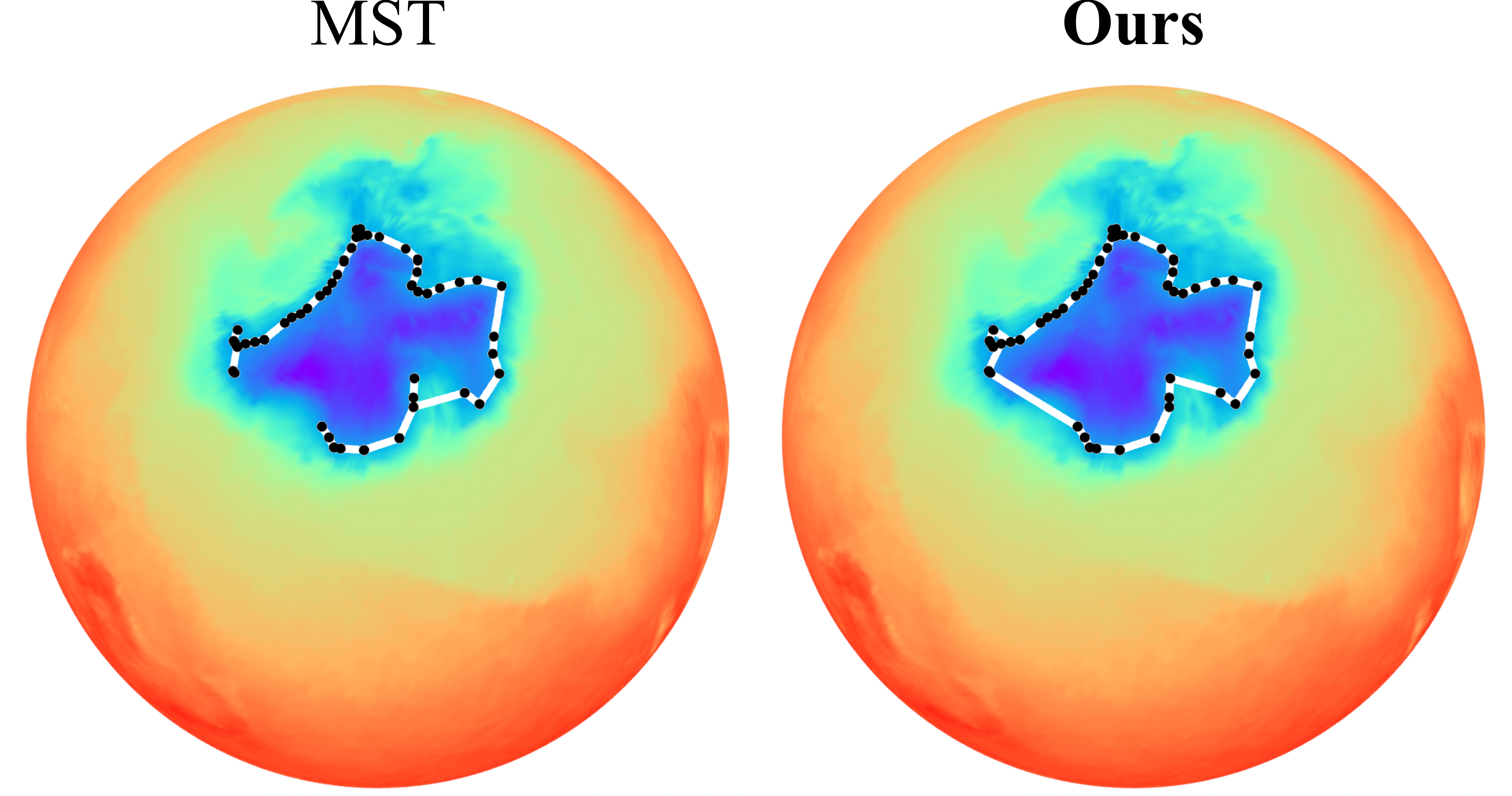}
    \caption{Samples (black) with approximately the same temperatures connected using the MST algorithm (left) and our method (right) to form a contour line (white) for a region with a temperature of 230$\pm$0.3K on the globe, for a single timestep.
        }
    \label{fig:temp_isoling}
\end{figure}

In \Cref{fig:temp_isoling}, we use the global temperature values recorded on 31$^\text{st}$ of March 2024, at midnight, and use a subset of the values (a quarter of the initial array, by skipping every second and fourth row) to minimize the amount of data. The texture of the mesh represents the ground truth temperatures. We then extract the points with a temperature of 230$\pm$0.3K and connect them to obtain the corresponding contour line, only requiring 63 samples to reconstruct the isoline. In contrast, the MST algorithm is unable to reconstruct the curve due to the non-uniformity of the samples. The data we used is based on data and products of the European Centre for Medium-Range Weather Forecasts \href{www.ecmwf.int}{(ECMWF)}, under CC BY 4.0. license.

%% file: 05-conclusions/main.tex
We have generalized the sampling requirements of closed curves from the Euclidean plane $\Reals^{2}$ to Riemannian manifold domains of arbitrary dimension. By taking inspiration from the literature on planar curve reconstruction, we designed a new method for recovering the original connectivity from a sparse sampling by biasing an instance of the traveling salesman problem with a coarse graph, which exhibits strong theoretical guarantees. We have proved the effectiveness of our method by testing it in different scenarios and various applications involving real-world data, showing that it outperforms the existing previous solution and that it provides a correct result, while in many cases the state-of-the-art method fails.

As for future directions, we intend to explore possible relaxations of the starting graph to offer stronger theoretical guarantees even under adversarial samplings. Furthermore, we notice that when the sampling conditions satisfy our constraints, the original reconstruction forms a Hamiltonian cycle inside the SIGDV graph. \revised{Since this path is not guaranteed to be the shortest tour, an algorithm for the TSP could be a sub-optimal strategy and} more sophisticated techniques for identifying such a cycle could be worth future investigation. Another possible avenue of future development \revised{could involve a more compelling approach for handling multiple closed loops and the exploration of techniques for} dealing with more general classes of curves, as these could provide an important tool in pattern extraction, where not all elements are closed smooth curves. Finally, as the reconstruction of curves in non-Euclidean domains is a barely explored topic, we identify the absence of a unified validation benchmark for quantitative analyses and we intend to fill this gap in the future.

%% file: a0-proofs/main.tex
In this appendix, we provide the proofs of all our claims.

\begin{customlem}{\ref{lem:anytopodisk}}
    Let $\C \subset \M$ be a closed curve on the manifold $\M$. For every point $p \in \M$ and every positive real value $r \leq i_{\M}(p)$, if the $r$-ball $\rball{p}{r}$ centered at $p$ contains at least two points of $\C$, then then either $\rball{p}{r}\cap\C$ is a topological 1-disk, or $\rball{p}{r}$ contains a point of $\Gamma(\C)$, or both.

    \begin{proof}
        \revised{%
        We denote the intersection between the $r$-ball and the curve with $\F = \rball{p}{r} \cap \C$. 
        
        If $\F$ is a topological 1-disk, there is nothing to prove. 

        If $\F \neq \C$ and $\F$ is not a topological 1-disk, then $\F$ must contain at least two connected components. Let $q_1 \in \F$ be the closest point in $\F$ to $p$, and let $f_1$ be the connected component of $\F$ containing $q_1$. If $q_1$ is not unique, then $p \in \Gamma(\C)$ and the proof is complete.
        Otherwise, let $f_2 \neq f_1$ be a second closest connected component of $\F$ to $p$, and let $q_2 \in f_2$ be the point in $f_2$ closest to $p$. Let $\delta_i(x) = \inf_{y \in f_i}d_{\M}(x, y)$ be the distance from each point $x$ to the connected component $f_i$. Let us consider the geodesic shortest path $\gamma_p^{q_2}$, which is contained in $\rball{p}{r}$ by definition of $r$-ball. We know that at point $p$ it holds that $\delta_1(p) < \delta_2(p)$ and at $q_2$ it holds that $\delta_1(q_2) > \delta_2(q_2)$. Since the distance functions change continuously, there must be some point $x$ along the path $\gamma_p^{q_2}$ such that $\delta_1(x) = \delta_2(x)$, where the closest connected component of $\F$ to $x$ changes. If $x \notin \Gamma(\C)$, then there must be another point $q_3 \in \C$ such that $q_3 \notin f_1, f_2$ and $d_{\M}(x, q_3) < \delta_2(x) \leq d_{\M}(x, q_2)$. 
        Using the triangle inequality, we get that $d_{\M}(p, q_3) < d_{\M}(p, x) + d_{\M}(x, q_3)$. Hence, we get that $d_{\M}(p, q_3) < d_{\M}(p, x) + d_{\M}(x, q_3) < d_{\M}(p, x) + d_{\M}(x, q_2)$. Using the fact that $x$ lies on the geodesic shortest path between $p$ and $q_2$, we have $d_{\M}(p, x) + d_{\M}(x, q_2) = d_{\M}(p, q_2)$, which implies that $d_{\M}(p, q_3) < d_{\M}(p, q_2)$, which contradicts our assumption that $f_2$, which contains $q_2$, is the second closest connected component (since $q_3\notin f_1, f_2$). Thus, $x \in \rball{p}{r}$ must belong to the medial axis $\Gamma(\C)$.

        If $\F = \C$, then the curve $\C$ is entirely contained in $\rball{p}{r}$. Let $\delta_{\C}(x) = \min_{y \in \C}d_{\M}(x, y)$ be the minimal distance from each point to the curve. We consider the $\ell$-neighborhood $T_{\ell}(\C) = \{ x \in \M : \delta_{\C}(x) < \ell \}$ of the curve $\C$ and the restriction of the neighborhood $N_{\ell}(\C) = \{ x \in \rball{p}{r} : \delta_{\C}(x) < \ell \}$ to the $r$-ball $\rball{p}{r}$. Since $\rball{p}{r}$ is an open set, for small values of $\ell$, the $\ell$-neighborhood $T_{\ell}(\C)$ is topologically equivalent to a $d$-dimensional solid torus and is entirely contained within $\rball{p}{r}$. Thus $N_{\ell}(\C) = T_{\ell}(\C)$ has genus $1$. On the other hand, $\rball{p}{r}$ has finite size, and thus for values of $\ell$ large enough $N_{\ell}(\C)$ covers the entire $r$-ball and hence, $N_{\ell}(\C) = \rball{p}{r}$. Since $r \leq i_{\M}(p)$, the $r$-ball $\rball{p}{r}$ is topologically equivalent to a $d$-dimensional Euclidean ball (whose genus is $0$), and so must be $N_{\ell}(\C)$. Since the curve $\C$ is immutable during the growing process of $N_{\ell}(\C)$, and no elements are removed while increasing $\ell$, the genus decrease of $N_{\ell}(\C)$ cannot happen via cutting the solid torus. Thus, for some $\ell$ there must be a non-contractible curve along the boundary of $N_{\ell}(\C)$ that collapses in a single point $q \in \rball{p}{r}$ to achieve a genus decrease. This means that $q$ must be at a distance $\ell$ from at least two different points on the curve, and thus, must be a point on the medial axis.
        }
            
        
    \end{proof}
\end{customlem}

\begin{customcor}{\ref{cor:curvetopodisk}}
    For every point $p \in \C$, and for $r \leq \glfs{p}$, the ball $\rball{p}{r}$ intersects $\C$ in a topological 1-disk.

    \begin{proof}
        Let $\rball{p}{r}$ be a $r$-ball centered at $p$, for some $r \in \Reals^+$, that does not intersect $\C$ in a topological 1-disk. If that is the case, then either $r > i_{\M} > \glfs{p}$ or $\rball{p}{r}$ contains at least one point $m$ of the medial axis (by \Cref{lem:anytopodisk}). This would lead to $r > d_{\M}(p, m) > \lfs{p} > \glfs{p}$.
        
    \end{proof} 
\end{customcor}

\begin{customprp}{\ref{prp:nadj1reach}}
    Let $S \subset \C$ be a sampling of the curve. For every point $p \in \C$, let $s_0, s_1 \in S$ be the samples such that the interval $I = (s_0, s_1)$ is the smallest open interval between samples that contains $p$. If there exists a point $q \in\C$ not belonging the the closure of $I$ that is closer to $p$ than both $s_0$ and $s_1$ are to $p$, then $d_{\M}(p, q) \geq \glfs{p}$.

    \begin{proof}
        Let us denote $\delta = d_{\M}(p, q)$, and consider the $r$-ball $\rball{p}{\delta}$ of size $\delta$ and centered at $p$.
        
        Since $s_0$ and $s_1$ lie outside $\rball{p}{\delta}$, $p$ lies inside $\rball{p}{\delta}$, and $\rball{p}{\delta}$ touches $q$, then either the intersection $\F = \C \cap \rball{p}{\delta}$ has two connected components, or $\C$ is tangent to $\rball{p}{\delta}$ at $q$.
        
        If $\F$ has two connected components, then by \Cref{cor:curvetopodisk} $\delta > \glfs{p}$.
        
        If $\C$ is is tangent to $\rball{p}{\delta}$ at $q$, then for every $\varepsilon > 0$, the ball $\rball{p}{\delta + \varepsilon}$ would intersect $\C$ in two connected components, meaning $\delta + \varepsilon > \glfs{p}$. Hence, $\delta \geq \glfs{p}$.
    \end{proof}
\end{customprp}

\begin{customcor}{\ref{cor:adjsample}}
    Let $S \subset \C$ be a sampling of the curve, and $s_0, s_1 \in S$ be two adjacent samples defining an interval $I = [s_0, s_1] \subset \C$. If $S$ is a $\rho$-sampling with $\rho < 1$, then for every point $p \in I$, the closest sample to $p$ is either $s_0$ or $s_1$.

    \begin{proof}
        Assume the closest sample to $p$ is some $s_i \neq s_0, s_1$, and let us denote $\delta = \min(d_{\M}(p, s_0), d_{\M}(p, s_1))$. Then $\delta > d_{\M}(p, s_i) \geq \glfs{p} \geq \greach{I}$. But by definition of $\rho$-sampling, $\delta < \rho \greach{I}$, which contradicts $\delta > \greach{I}$ for $\rho < 1$.
    \end{proof}
\end{customcor}

\begin{customprp}{\ref{prp:adj2reach}}
    Let $S \subset \C$ be a $\rho$-sampling, with $\rho < 1$. For any two consecutive samples $s_0, s_1$, let $I = [s_0, s_1]$ be the interval between them. Then we have $d_{\M}(s_0, s_1) < 2 \greach{I}$.

    \begin{proof}
        Consider a point $p$ in the interval $I = [s_0, s_1]$ such that $d_{\M}(s_0, p) = d_{\M}(s_1, p)$. That point must be at $d \geq \frac{1}{2}d_{\M}(s_0, s_1)$ from $s_0$, and by definition of $\rho$-sampling with $\rho < 1$, $d < \greach{I} \leq \glfs{p}$. By \Cref{cor:adjsample}, the closest samples to $p$ must be $s_0$ and $s_1$, and we have $d_{\M}(s_0, s_1) \leq 2d < 2\ \greach{I}$.
    \end{proof}
\end{customprp}

\begin{customlem}{\ref{lem:dvedge}}
    Let $s_0, s_1 \in \C$ be consecutive samples from a $\rho$-sampling $S \in C$, with $\rho < 1$. The edge $e(s_0, s_1)$ belongs to the dual Voronoi graph of $S$.

    \begin{proof}
        Let $I = [s_0, s_1] \subset \C$ be the interval of $\C$ connecting samples $s_0$ and $s_1$, and let $p \in I$ be a point such that $d_{\M}(s_0, p) = d_{\M}(s_1, p)$. We consider the $r$-ball $\rball{p}{r}$ centered at $p$, where $r = d_{\M}(p, s_i)$ is the distance from $p$ to its closest sample $s_i$.
        
        By \Cref{cor:adjsample}, $s_i$ must be either $s_0$ or $s_1$, yielding $r = d_{\M}(s_0, p) = d_{\M}(s_1, p)$. Then the boundary between the Voronoi cells of $s_0$ and $s_1$ is not empty (as it contains at least $p$), and the edge $e(s_0, s_1)$ is the dual of their shared Voronoi boundary.
    \end{proof}
\end{customlem}

\begin{customlem}{\ref{lem:sigedge}}
    Let $s_0, s_1, s_2, s_3 \in \C$ be consecutive samples from a $\rho,u$-sampling $S \in C$, with $\rho < 1$ and $u < 2$. The edge $e(s_1, s_2)$ belongs to the Spheres-of-Influence graph of $S$.

    \begin{proof}
        We split the proof into four cases.

        Case 1: If $s_1$ is the nearest neighbor of $s_2$, or vice versa, then the edge $e$ trivially belongs to SIG.
        
        Case 2: If $s_0, s_3$ are the nearest neighbors of $s_1$ and $s_2$ respectively, the non-uniformity ratio $u < 2$ imposes $d_{\M}(s_0, s_1), d_{\M}(s_2, s_3) > \frac{1}{2}d_{\M}(s_1, s_2)$, meaning that $d_{\M}(s_0, s_1) + d_{\M}(s_2, s_3) > d_{\M}(s_1, s_2)$. Then the edge $e$ belongs to SIG by definition.
        
        Case 3: Now, suppose neither $s_0$ nor $s_2$ is a nearest neighbor for $s_1$, but $s_3$ is a nearest neighbor for $s_2$. Then, the distance from $s_2$ to its nearest neighbor is $d_2 = d_{\M}(s_2, s_3)$, and let $d_1 = d_{\M}(s_1, s_i)$ be the distance from $s_1$ to its nearest neighbor $s_i$. If $d_1 + d_2 \geq d_{\M}(s_1, s_2)$, the edge belongs to SIG by definition. 
        
        Otherwise, we must have $d_1 < d_{\M}(s_1, s_2) - d_2$. Because of the non-uniformity ratio $u < 2$, we know that $\frac{1}{2}d_{\M}(s_1, s_2) < d_2 < 2d_{\M}(s_1, s_2)$, and hence $d_1 < \frac{1}{2}d_{\M}(s_1, s_2)$. By \Cref{prp:adj2reach}, $d_1 < \greach{[s_1, s_2]} \leq \glfs{s_1}$. However, since $s_i$ is not adjacent to $s_1$, by \Cref{prp:nadj1reach} we have $d_1 \geq \glfs{s_1}$, which contradicts the above $d_1 < \glfs{s_1}$.
        
        Case 4: Finally, suppose neither $s_0$ nor $s_2$ is a nearest neighbor for $s_1$, and neither $s_1$ nor $s_3$ is a nearest neighbor for $s_2$. If that is the case, a sample $s_i$ must lie at distance $d_1 = d_{\M}(s_i, s_1)$, and \Cref{prp:nadj1reach} requires $d_1 \geq \glfs{s_1}$. We also must have a sample $s_j$ lie at distance $d_2 = d_{\M}(s_2, s_j)$ from $s_2$, which is the nearest neighbor of $s_2$, and \Cref{prp:nadj1reach} requires $d_2 \geq \glfs{s_2}$. If $d_1 + d_2 \geq d_{\M}(s_1, s_2)$, then the edge $e$ belongs to SIG by definition.
        Suppose then $d_1 + d_2 < d_{\M}(s_1, s_2)$, and let us denote $r_{1, 2} = \min(\glfs{s_1}, \glfs{s_2})$. \Cref{prp:adj2reach} requires $d_1 + d_2 < d_{\M}(s_1, s_2) < 2 r_{1, 2}$. Since $d_1 \geq \glfs{s_1} \geq r_{1, 2}$, we have $d_2 < r_{1, 2}$, which contradicts $d_2 \geq \glfs{s_2} \geq r_{1, 2}$.
    \end{proof}
\end{customlem}

\begin{customthm}{\ref{thm:sigdvedge}}
    Let $\M$ be a $d$-dimensional Riemannian manifold, possibly with boundary $\partial\M$, and equipped with a geodesic distance $d_{\M} : \M\times\M \to  \Reals$. Let $\C \subset \M$ be a curve on the manifold $\M$, and $S = \{ s_1, \cdots, s_k \} \subset \C$ be a $\rho,u$-sampling of $\C$, with $\rho < 1$ and $u < 2$. The edge connecting any pair of consecutive samples is part of the SIGDV of $S$.
    
    \begin{proof}
        By \Cref{lem:dvedge} the edge $e$ belongs to the dual Voronoi connectivity, and by \Cref{lem:sigedge} the edge $e$ belongs to SIG. Then the edge also belongs to their intersection SIGDV.
    \end{proof}
\end{customthm}

\begin{customprp}{\ref{prp:so3r4}}
    Given a set of rotations $R = \{ r_i \}_{i = 1}^{k}$ in $\so{3}$ and a set of quaternions $Q = \{ q_i \}_{i = 1}^{k}$ in $\mathbb{H}$ such that $q_i$ represents the rotation $r_i$, then the Voronoi decomposition $\mathrm{Vor}(R, d_{\so{3}})$ of $\so{3}$ w.r.t. the rotations $R$ can be obtained as
    \begin{equation}
        \mathrm{Vor}(R, d_{\so{3}})
        =
        \mathrm{Vor}(Q, d_{\Realsn{4}})
        \cap
        \so{3}\,.
    \end{equation}

    \begin{proof}
        Let $r_p, r_q \in \so{3}$ be two rotations, and let $p = (p_w, p_x, p_y, p_z), q = (q_w, q_x, q_y, q_z) \in \mathbb{H}$ be the two quaternions representing them. Let $q^{*}$ be the conjugate of $q$. We know that the angular distance between $r_p$ and $r_q$ can be expressed as the angle of rotation of the quaternion $t = (t_w, t_x, t_y, t_z) = pq^{*}$, which in turn is given by $2\arccos(t_w) = 2\arccos(\langle p, q \rangle)$.

        We also know that the squared Euclidean distance between $p$ and $q$ is given by $\| p - q \|^2 = \| p \|^2 + \| q \|^2 - 2 \langle p, q \rangle$.
        Since the quaternions express rotations, they have unitary norm, meaning $\| p - q \|^2 = 2 - 2 \langle p, q \rangle$. Thus, the angular distance between $r_p$ and $r_q$ can be expressed as
        \begin{equation*}
            d_{\so{3}}(r_p, r_q)
            =
            2 \arccos(\langle p, q \rangle)
            =
            2 \arccos\left( 1 - \frac{\|p - q \|^2}{2} \right)\,.
        \end{equation*}
        
        Since $\langle p, q \rangle$ is bounded by $-1$ and $1$ ($\langle p, q \rangle = \|p\|\|q\|\cos(\theta) = \cos(\theta)$, as $p$ and $q$ are unitary), and since $\arccos(\cdot)$ is a decreasing function on $[-1, 1]$, then for any rotation $r_w \in \so{3}$ and its corresponding quaternion $w \in \mathbb{H}$ we get that $d_{\so{3}}(r_p, r_w) < d_{\so{3}}(r_q, r_w) \iff \langle p, w \rangle > \langle q, w \rangle$, but we also know that $\langle p, w \rangle > \langle q, w \rangle \iff \|p - w \|^2 < \| q - w \|^2$. Thus, it must be true that 
        \begin{equation*}
            d_{\so{3}}(r_p, r_w) < d_{\so{3}}(r_q, r_w)
            \quad\iff\quad
            \|p - w \| < \| q - w \|\,.
        \end{equation*}
        
        This proves that the distance functions $d_{\so{3}}$ and $d_{\Reals{4}}$ preserve the distance relationships between rotations in $\so{3}$, meaning that $\mathrm{Vor}(R, d_{\so{3}}) = \mathrm{Vor}(Q, d_{\Realsn{4}}) \cap \so{3}$.
    \end{proof}
\end{customprp}